\documentclass{aa}

\usepackage{graphicx}
\usepackage{txfonts}
\usepackage{hyperref}
\usepackage{lscape}

\usepackage{natbib}
\bibpunct{(}{)}{;}{a}{}{,} 

\begin{document}

	\title{Chemical abundances and deviations from the solar S/O ratio in the gas-phase ISM of galaxies based on infrared emission lines}
	\titlerunning{Abundances and deviations from solar S/O based on IR lines}

	\author{Borja Pérez-Díaz\inst{\ref{inst1}}
		\and
		Enrique Pérez-Montero\inst{\ref{inst1}} \and Juan A. Fernández-Ontiveros\inst{\ref{inst2}} \and José M. Vílchez\inst{\ref{inst1}} \and Antonio Hernán-Caballero\inst{\ref{inst2}} \and Ricardo Amorín\inst{\ref{inst2}}}

	\institute{Instituto de Astrofísica de Andalucía (IAA-CSIC), Glorieta de la Astronomía s/n, 18008 Granada, Spain\label{inst1} \\\email{bperez@iaa.es}\and Centro de Estudios de Física del Cosmos de Aragón (CEFCA), Unidad Asociada al CSIC, Plaza San Juan 1, E--44001 Teruel, Spain\label{inst2}}

	\date{Received MONTH DAY, YEAR; accepted MONTH DAY, YEAR}



\abstract
{The infrared (IR) range is extremely useful in the context of chemical abundance studies of the gas-phase interstellar medium (ISM) due to the large variety of ionic species traced in this regime, the negligible effects from dust attenuation or temperature stratification, and the amount of data that has been and will be released in the coming years.}
{Taking advantage of available IR emission lines, we analysed the chemical content of the gas-phase ISM in a sample of 131 Star-Forming Galaxies (SFGs) and 73 Active Galactic Nuclei (AGNs). Particularly, we derived the chemical content via their total oxygen abundance in combination with nitrogen and sulfur abundances, and with the ionisation parameter.}
{We used a new version of the code \textsc{HII-CHI-Mistry-IR} v3.1 which allows us to estimate log(N/O), 12+log(O/H), log(U), and, for the first time, 12+log(S/H) from IR emission lines, which can be applied to both SFGs and AGNs. We tested that the estimations from this new version, that only considers sulfur lines for the derivation of sulfur abundances, are compatible with previous studies.}
{While most of the SFGs and AGNs show solar log(N/O) abundances, we found a large spread in the log(S/O) relative abundances. Specifically, we found extremely low log(S/O) values (1/10th solar) in some SFGs and AGNs with solar-like oxygen abundances. This result warns against the use of optical and IR sulfur emission lines to estimate oxygen abundances when no prior estimation of log(S/O) is provided.}
{}


\maketitle



\section{Introduction}
\label{sec1}
Emission lines measured in the gas phase of the interstellar medium (ISM) in galaxies are key to infer their physical and chemical properties. Among these emission lines, collisionally excited lines (CELs) have been widely used for this purpose in several spectral ranges, such as in the optical \citep[e.g.][]{Lequeux_1979, Garnett_1987, Contini_2001, Perez-Montero_2014, Curti_2017}, the ultraviolet \citep[UV; e.g.][]{Erb_2010, Dors_2014, Perez-Montero_2017}, and the infrared \citep[IR; e.g.][]{Nagao_2011, Pereira-Santaella_2017, Peng_2021, Fernandez-Ontiveros_2021, Perez-Diaz_2022}. As the primordial ISM metal content after the Big Bang nucleosynthesis is well constrained \citep{Cyburt_2016}, any subsequent deviation from these initial chemical conditions must be attributed to stellar nucleosynthesis, whose products are ejected into the ISM in the late stages of stellar evolution. Therefore, the analysis of the metal content of the ISM is fundamental to understanding the impact of dissipative baryonic processes in galaxy evolution.

Over decades, several techniques based on CELs have been developed and improved to infer the metal content of the ISM in Star-Forming Galaxies \citep[SFGs; see][for a review on the topic]{Maiolino_2019}. Moreover, in recent years, similar techniques have also started to be applied for the study of the chemical content of the ISM in the Narrow Line Region (NLR) of Active Galactic Nuclei (AGNs), accounting for the corresponding differences in the sources ionising  the surrounding gas \citep[e.g.][]{Contini_2001, Dors_2015, Perez-Montero_2019, Perez-Diaz_2021}.

Most of the studies devoted to the analysis of chemical abundances of the gas-phase ISM using CELs are focused on the oxygen content (12+log(O/H)), due to several reasons: \textit{i)} O is the most abundant metal by mass in the gas-phase ISM \citep[\ensuremath{\sim}55\%,][]{Peimbert_2007}, and is therefore a good proxy for the total metallicity ($Z$); and, \textit{ii)} its abundance can be derived more easily than other elements due to the presence of strong CELs in the optical, IR and UV spectral ranges. Additionally, some authors have also analysed the nitrogen-to-oxygen abundance ratio \citep[log(N/O); e.g.][]{Vila-Costas_1993, Perez-Montero_2009, Amorin_2010, Andrews_2013, Peng_2021, Fernandez-Ontiveros_2021, Perez-Diaz_2022}. Nitrogen N can be produced by massive stars via a primary channel --\,leading to an almost constant N/O ratio\,-- but also through a secondary channel in the high-metallicity regime, due to CNO cycles in intermediate mass stars which eject it into the ISM after a certain time delay \citep[e.g.][]{Henry_2000}. Thus, the study of N/O using nitrogen emission lines can provide complementary information on the evolution of the chemical content of the ISM. While studies of N/O from both optical and IR emission lines can be performed, in the UV other secondary elements --\,such as carbon (C)\,-- are studied instead, due to the presence of strong emission lines in this range, although this study is also possible by using optical recombination lines (RLs) \citep[e.g.][]{Toribio_2017, Martinez-Delgado_2022}, with the disadvantage that these emission lines are much fainter than CELs \citep{Esteban_2009}.

So far, very few works have studied in a statistically significant sample of galaxies the sulfur (S) content in the gas-phase ISM. Instead, the assumption of an universal S/O ratio has been used to propose sulfur emission-lines as tracers of the total oxygen metallicity. For instance, \citet{Vilchez_1996} were pioneers in defining the sulfur abundance parameter S{\ensuremath{_{23}}}:
\begin{equation}
\label{S23} \mathrm{S}_{23} = \frac{ \mathrm{I} \left(  \left[ \mathrm{\ion{S}{ii}} \right]\lambda \lambda 6717,6731 \right) +  \mathrm{I} \left( \left[ \mathrm{\ion{S}{iii}} \right] \lambda \lambda 9069,9532 \right) }{ \mathrm{I} \left( \mathrm{H}_{\beta } \right) }
\end{equation}

Later on, \citep{Diaz_2000} provided the first calibration to directly estimate the oxygen content from S{\ensuremath{_{23}}}, improving the determination of the chemical content of the gas-phase ISM, counteracting the ambiguity of the equivalent oxygen parameter R$_{23}$ \citep{Pagel_1979}. As shown by these authors, and after further improvements on the calibration of this estimator \citep{Perez-Montero_2005}, two advantages arise from using sulfur emission lines instead of oxygen (i.e. R$_{23}$): \textit{i)}, S{\ensuremath{_{23}}} is mainly single-valued in most of the metal abundance range; and, \textit{ii)} sulfur lines are less affected by extinction than oxygen emission lines although [\ion{S}{iii}] lines at 9069\ensuremath{\AA} and 9531\ensuremath{\AA} can suffer from telluric absorptions \citep[e.g.][]{Noll_2012}.

However, the use of S$_{23}$ to estimate 12+log(O/H) directly implies that the ratio between S and O (log(S/O)) remains constant. Indeed, while S and O are both produced in the nucleosynthesis of massive stars, their yields are expected to also behave similarly, supporting the previous idea of a constant S/O ratio. Unfortunately this assumption has not been firmly established, and only a few works have analysed the sulfur content in the ISM \citep[e.g.][]{Kehrig_2006, Perez-Montero_2006, Diaz_2007, Berg_2020, Diaz_2022, Dors_2023} as compared to the large number of studies on the oxygen content. Therefore, further observational constrains are required to validate the assumption of a universal solar S/O ratio. For instance, while \citet{Berg_2020} show that most {\sc H\,ii} regions in their sample are consistent with a S/O solar ratio, \citet{Diaz_2022} found strong deviations from that, especially in the low-metallicity regime (12+log(O/H) < 8.1). In this regard, it is important to note that at low metallicities a higher ionisation degree of the gas-phase ISM is expected, so higher ionic species (such as S$^{+3}$) contribute more to the total budget of the sulfur content and thus the uncertainty due to the application of the ICF to optical lines is higher. Moreover, \citet{Dors_2023} also found a few sources among their AGN sample with S/O ratios in some galaxies far from the solar value. Several attempts have been also made to directly calibrate the S23 parameter with the total sulphur abundance \citep{Perez-Montero_2006, Diaz_2022}. However, the collisional nature of the lines, involved in this parameter, which make them to be very dependent on the electron temperature and thus on the overall metal content of the gas, implies an additional dependence on the assumed S/O ratio.

Nevertheless, the above mentioned studies focus on the use of optical emission lines, leading to an inconclusive response whether these deviations originate intrinsically in the production of S and O, or due to a variety of of effects with diverse origins such as dust attenuation, contamination from diffuse ionised gas (DIG) or the effects from the assumed ionisation correction factors (ICFs), which severely affect the total sulphur abundances derived from the optical emission lines, as these do not cover the higher ionised S stages, such as S$*{3+}$. In this regard, the study of IR emission lines opens a new avenue to determine sulfur abundances, both in SFGs and AGNs, with key advantages over optical tracers. First of all, due to the atomic transitions involved in the ionic radiative process, the IR emission lines are much less affected by the electron temperature T{\ensuremath{_{e}}}, avoiding problems due to stratification or temperature fluctuations \citep[e.g.][]{Peimbert_1967, Stasinska_2005, Martinez-Delgado_2023, Jin_2023}. Secondly, the IR range allows the detection of highly ionised species such as S{\ensuremath{^{+3}}}, which are important for a more accurate determination of total elemental abundances, especially in AGNs. Thirdly, IR emission lines are almost unaffected by dust obscuration. Fourthly, ancillary data from observatories such as \textit{Infrared Space Observatory} (\textit{ISO}, covering the 2.4-197 $\mu$m range, \citealt{Kessler_1996}), the \textit{Spitzer Space Observatory} (5-39 $\mu $m, \citealt{Werner_2004}), the \textit{Herschel Space Observatory} (51-671 $\mu $m, \citealt{Pilbratt_2010}) and the \textit{Stratospheric Observatory for Infrared Astronomy} (\textit{SOFIA}, covering the 50-205 $\mu $m range, \citealt{Fischer_2018}) as well as brand-new missions such as the \textit{James Webb Space Telescope} (\textit{JWST}, which is covering the 4.9-28.9$\mu$m range with the Mid-InfraRed Instrument MIRI, \citealt{Rieke_2015, Wright_2015}) and upcoming facilities such as the Mid-infrared Extremely large Telecope Imager and Spectrograph (METIS, covering the N-band centered at 10$\mu$m, \citealt{Brandl_2021}).
\begin{figure*}
	\centering
	\includegraphics[width=0.7\hsize]{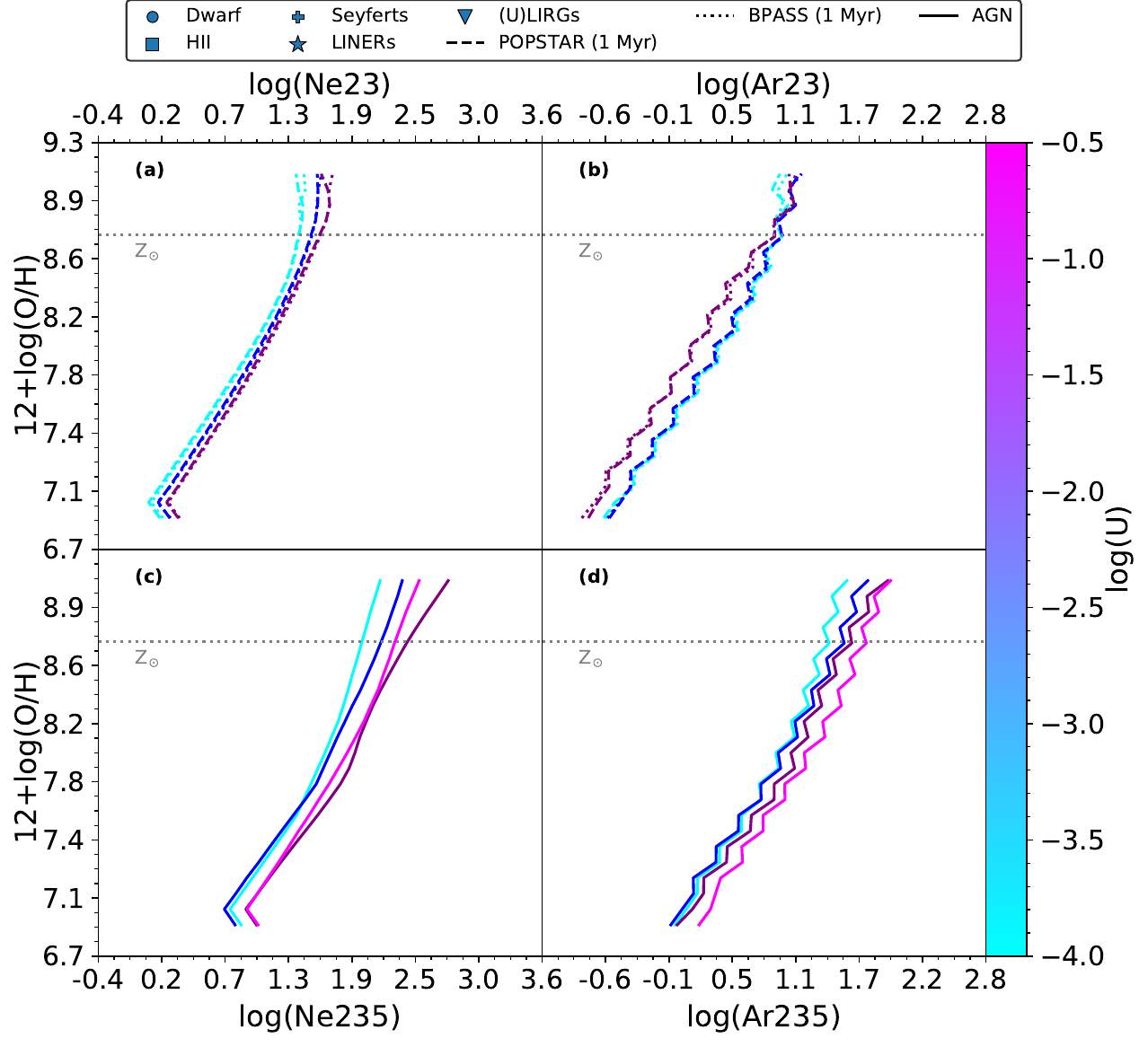}
	\caption{Performance of estimators based on neon (a) and (c) and argon (b) and (d) emission lines. Panels (a) and (b) show the behavior of estimators Ne23 and Ar23 for SFG models, with plus symbols presenting values from our SFG sample. Panels (c) and (d) show the behavior for estimators Ne235 and Ar235 for AGNs. SF models presented by \citet{Fernandez-Ontiveros_2021} are shown as dashed lines, while AGN models presented by \citet{Perez-Diaz_2022} are shown as continuous lines.}
	\label{ar_estimators}
\end{figure*}

In this work, we compiled a sample of SFGs and AGNs with IR spectroscopic observations to derive their chemical abundances from the IR emission lines, following the methodology used in \citet{Fernandez-Ontiveros_2021} for SFGs and in \citet{Perez-Diaz_2022} for AGNs, which includes an independent estimation of log(N/O), 12+log(O/H) and 12+log(S/H). The paper is organised as follows. Section \ref{sec2} provides information on the sample selection as well as on the methodlogy followed through this work. The main results of this study are shown in Section \ref{sec3} and discussed in Section \ref{sec4}. Finally, we summarise our conclusion in Section \ref{sec5}.

\section{Sample and methodology}
\label{sec2}

For this work we compiled one of the largest sample of galaxies with IR spectroscopic observations combining catalogs from {\it Spitzer}, {\it Herschel}, Akari and SOFIA. Particularly, we compiled the following catalogs: the dwarf galaxies sample from \citet{Cormier_2015} observed with {\it Spitzer} and {\it Herschel}; the Infrared Database of Extragalactic Observables from Spitzer\footnote{http://ideos.astro.cornell.edu/} \citep[IDEOS][]{Hernan-Caballero_2016, Spoon_2022} from {\it Spitzer} archive; the AGN and HII samples from \citet{Fernandez-Ontiveros_2016} and \citet{Spinoglio_2022} combining {\it Spitzer}, {\it Herschel} and SOFIA; and the (U)LIRG catalog from \citet{Imanishi_2010} observed with Akari.

\subsection{SFG sample}
Our sample of SFGs is composed by objects in two main sources. The data of the first subsample were directly taken from \citet{Fernandez-Ontiveros_2021}, where they compiled a sample of 65 galaxies (30 dwarf galaxies, 22 \textsc{H\,ii} regions and 13 (U)LIRGs) with IR spectroscopic observations showing star-formation dominated emission ([NeV]/[NeII] < 0.15). Additionally, we compiled another sample of galaxies from the \textsc{IDEOS} catalog \citep{Hernan-Caballero_2016, Spoon_2022}. As described in \citet{Perez-Diaz_2023}, the sample consists of 66 Ultra-Luminous Infrared Galaxies (ULIRGs) showing star-forming dominated activity from both their [Ne{\sc v}]/[Ne{\sc ii}] ratio (< 0.15) and from the equivalent width of the PAH feature at 6.2{\ensuremath{\mu}}m (EQW(PAH\ensuremath{_{6.2\mu m}}) > 0.06 {\ensuremath{\mu}}m). While for this last sample we compiled IR emission lines from [S{\sc iv}]\ensuremath{\lambda}10\ensuremath{\mu}m to [S{\sc iii}]\ensuremath{\lambda}33\ensuremath{\mu}m from IDEOS \citep{Spoon_2022} and measurements of H{\sc I}\ensuremath{\lambda}4.05\ensuremath{\mu}m from Akari/IRC observations \citep[2.5-5{\ensuremath{\mu}}m,][]{Imanishi_2010}, only the first sample from \citet{Fernandez-Ontiveros_2021} presents measurements from far-IR emission lines such as [O{\sc iii}]\ensuremath{\lambda}52\ensuremath{\mu}m, [N{\sc iii}]\ensuremath{\lambda}57\ensuremath{\mu}m and [O{\sc iii}]\ensuremath{\lambda}88\ensuremath{\mu}m, which are key to estimate N/O.

\subsection{AGN sample}
We compiled our AGN sample from \citet{Perez-Diaz_2022}, who analysed 58 AGNs with available IR spectroscopic observations, including 17 Seyfert 1 nuclei (Sy1), 14 Seyfert nuclei with hidden broad lines in the polarised spectrum (Sy1h), 12 Seyfert 2 nuclei (Sy2), 12 (U)LIRGs and 3 LINERs. Additionally, we included 15 quasars from the IDEOS catalog (up to redshift {\ensuremath{\sim}} 0.74), for which we also added measurements, when possible, of the hydrogen recombination line H{\sc i} Brackett \ensuremath{\alpha} line, from their Akari/IRC observations \citep[2.5-5{\ensuremath{\mu}}m,][]{Imanishi_2010}. We measured fluxes in the {H{\sc i}$\lambda$4.05\ensuremath{\mu}m line by fitting the restframe [3.8--4.3\ensuremath{\mu}m] range with a model that assumes a 2$^{nd}$ order polynomial for the continuum and a Gaussian profile for the line, with the line width corresponding to the instrumental resolution of the Akari or Spitzer/IRS spectrum at that wavelength.\footnote{The spectral resolution of Akari is $R$$\sim$100 \citep{Kim_2015}. For Spitzer/IRS, it is $\Delta \lambda$ = 0.06 and 0.12 \ensuremath{\mu}m for the SL2 (5.15-7.5\ensuremath{\mu}m)and SL1 (7.5-14\ensuremath{\mu}m) modules, respectively \citep{Spoon_2022}.}. Both samples show strong AGN emission as shown by their [Ne{\sc v}]/[Ne{\sc ii}] (\ensuremath{\gg} 0.15) ratio.

\subsection{\textsc{HII-CHI-Mistry-IR}}

To derive chemical abundances from IR emission lines in our sample we used the code \textsc{HII-CHI-Mistry-IR} (hereinafter, \textsc{HCm-IR}) v3.1, originally developed by \citet{Perez-Montero_2014} for the optical emission-lines, and later extended to IR emission lines by \citet{Fernandez-Ontiveros_2021} for SFGs and \citet{Perez-Diaz_2022} for AGNs. This code basically performs a bayesian-like comparison between a set of observed emission-line flux ratios sensitive to quantities such as total oxygen abundance, nitrogen-to-oxygen ratio, or the ionisation parameter, with the predictions from large grids of photoionisation models to provide the most probable values of these quantities and their corresponding uncertainties.

The version 3.1 of the code for the IR\footnote{The code is publicly available at \url{http://home.iaa.csic.es/~epm/HII-CHI-mistry.html}.} presents two new features in relation to previous versions:
\begin{itemize}
\item The code now accepts as input the argon emission lines ([\ion{Ar}{ii}]{\ensuremath{\lambda}}7{\ensuremath{\mu}}m, [\ion{Ar}{iii}]{\ensuremath{\lambda}}9{\ensuremath{\mu}}m, [\ion{Ar}{v}]{\ensuremath{\lambda}}8{\ensuremath{\mu}}m and [\ion{Ar}{v}]{\ensuremath{\lambda}}13{\ensuremath{\mu}}m), which are used to construct estimators of metallicity and excitation, analogous to those based on neon emission lines already used in previous versions, i.e., Ne23 and Ne2Ne3 for SFGs \citep{Fernandez-Ontiveros_2021}:
\begin{equation}
\label{Ne23} \begin{aligned} \log \left( \mathrm{Ne23} \right) &  = & \log \left( \frac{ \mathrm{I} \left( \left[ \mathrm{\ion{Ne}{ii}} \right] _{12\mu m} \right) +  \mathrm{I} \left( \left[ \mathrm{\ion{Ne}{iii}} \right] _{15\mu m} \right) }{ \mathrm{I} \left( \mathrm{\ion{H}{i}}  _{i} \right) } \right)
\end{aligned},
\end{equation}
\begin{equation}
\label{Ne2Ne3} \log \left( \mathrm{Ne2Ne3} \right) = \log \left(  \frac{ \mathrm{I} \left( \left[ \mathrm{\ion{Ne}{ii}} \right] _{12\mu m} \right)  }{  \mathrm{I} \left( \left[ \mathrm{\ion{Ne}{iii}} \right] _{15\mu m} \right) }  \right),
\end{equation}
and Ne235 and Ne23Ne5 for AGNs \citep{Perez-Diaz_2022}:
\begin{equation}
\label{Ne235} \begin{aligned} \log \left( \mathrm{Ne235} \right) &  = & \log \left( \frac{ \mathrm{I} \left( \left[ \mathrm{\ion{Ne}{ii}} \right] _{12\mu m} \right) +  \mathrm{I} \left( \left[ \mathrm{\ion{Ne}{iii}} \right] _{15\mu m} \right) }{ \mathrm{I} \left( \mathrm{\ion{H}{i}}  _{i} \right) } + \right. \\ & & + \left. \frac{  \mathrm{I} \left( \left[ \mathrm{\ion{Ne}{v}} \right] _{14\mu m} \right) + \mathrm{I} \left( \left[ \mathrm{\ion{Ne}{v}} \right] _{24\mu m}  \right) }{  \mathrm{I} \left( \mathrm{\ion{H}{i}}  _{i} \right) }   \right)
\end{aligned},
\end{equation}
\begin{equation}
\label{Ne23Ne5} \log \left( \mathrm{Ne23Ne5} \right) = \log \left(  \frac{ \mathrm{I} \left( \left[ \mathrm{\ion{Ne}{ii}} \right] _{12\mu m} \right) + \mathrm{I} \left( \left[ \mathrm{\ion{Ne}{iii}} \right] _{15\mu m} \right)   }{  \mathrm{I} \left( \left[ \mathrm{\ion{Ne}{v}} \right] _{14\mu m} \right) +  \mathrm{I} \left( \left[ \mathrm{\ion{Ne}{v}} \right] _{24\mu m} \right) }  \right),
\end{equation} 
Following this approach, the new observables based on these IR argon lines can be defined as:
\begin{equation}
\label{Ar23} \begin{aligned} \log \left( \mathrm{Ar23} \right) &  = & \log \left( \frac{ \mathrm{I} \left( \left[ \mathrm{\ion{Ar}{ii}} \right] _{7\mu m} \right) +  \mathrm{I} \left( \left[ \mathrm{\ion{Ar}{iii}} \right] _{9\mu m} \right) }{ \mathrm{I} \left( \mathrm{\ion{H}{i}}  _{i} \right) } \right)
\end{aligned},
\end{equation}
\begin{equation}
\label{Ar2Ar3} \log \left( \mathrm{Ar2Ar3} \right) = \log \left(  \frac{ \mathrm{I} \left( \left[ \mathrm{\ion{Ar}{ii}} \right] _{7\mu m} \right)  }{  \mathrm{I} \left( \left[ \mathrm{\ion{Ar}{iii}} \right] _{9\mu m} \right) }  \right),
\end{equation}
\begin{equation}
\label{Ar235} \begin{aligned} \log \left( \mathrm{Ar235} \right) &  = & \log \left( \frac{ \mathrm{I} \left( \left[ \mathrm{\ion{Ar}{ii}} \right] _{7\mu m} \right) +  \mathrm{I} \left( \left[ \mathrm{\ion{Ar}{iii}} \right] _{9\mu m} \right) }{ \mathrm{I} \left( \mathrm{\ion{H}{i}}  _{i} \right) } + \right. \\ & & + \left. \frac{  \mathrm{I} \left( \left[ \mathrm{\ion{Ar}{v}} \right] _{8\mu m} \right) + \mathrm{I} \left( \left[ \mathrm{\ion{Ar}{v}} \right] _{13\mu m}  \right) }{  \mathrm{I} \left( \mathrm{\ion{H}{i}}  _{i} \right) }   \right)
\end{aligned},
\end{equation}
\begin{equation}
\label{Ar23Ar5} \log \left( \mathrm{Ar23Ar5} \right) = \log \left(  \frac{ \mathrm{I} \left( \left[ \mathrm{\ion{Ar}{ii}} \right] _{7\mu m} \right) + \mathrm{I} \left( \left[ \mathrm{\ion{Ar}{iii}} \right] _{9\mu m} \right)   }{  \mathrm{I} \left( \left[ \mathrm{\ion{Ar}{v}} \right] _{8\mu m} \right) +  \mathrm{I} \left( \left[ \mathrm{\ion{Ar}{v}} \right] _{13\mu m} \right) }  \right),
\end{equation}
with $\mathrm{\ion{H}{i}}  _{i}$ being one of the hydrogen lines that the code can take as input. The performance of these estimators is shown in Fig. \ref{ar_estimators} in comparison with their neon analogues. Argon emission lines offers a great opportunity for chemical abundance studies with JWST data, since they are located in a narrow IR window [7\ensuremath{\mu}m,13\ensuremath{\mu}m], argon is a non-depleted element whose nucleosynthesis leads to a yield which is similar to that for oxygen, and they include transitions from highly ionised species which will help to disentangle the power of the ionising source.

\item After the iteration\footnote{The code originally performs two consecutive iterations: during the first iteration, the grid of models in which N/O, O/H and U are free parameters is constrained by the estimation of N/O. During the second iteration, O/H and U are estimated from the constrained grid of models. With this new feature, the code performs, in parallel to the estimation of O/H and U, the estimation of S/H, i.e., the code estimates S/H from the grid constrained only by N/O.} performed by the code to constrain N/O, and parallel to the second iteration performed to estimate 12+log(O/H) and the ionisation parameter log($U$), now the code performs a new iteration to estimate 12+log(S/H). This is done using the estimators S34 and S3S4 \citep{Fernandez-Ontiveros_2021, Perez-Diaz_2022}, based on sulfur lines, without assuming the results from the oxygen or the ionisation estimations. In this way, the estimators based on sulfur emission-lines (S34 and S3S4) are no longer used for the estimation of 12+log(O/H) and log($U$) and, consequently, sulfur and oxygen abundances are derived independently from each others estimations.
\begin{equation}
\label{S34} \begin{aligned} \log \left( \mathrm{S34} \right) &  = & \log \left( \frac{ \mathrm{I} \left( \left[ \mathrm{\ion{S}{iii}} \right] _{18\mu m} \right) +  \mathrm{I} \left( \left[ \mathrm{\ion{S}{iii}} \right] _{33\mu m} \right) }{ \mathrm{I} \left( \mathrm{\ion{H}{i}}  _{i} \right) } + \right. \\ & & + \left. \frac{ \mathrm{I} \left( \left[ \mathrm{\ion{S}{iv}} \right] _{10\mu m}  \right) }{  \mathrm{I} \left( \mathrm{\ion{H}{i}}  _{i} \right) }   \right)
\end{aligned},
\end{equation}
\begin{equation}
\label{S3S4} \log \left( \mathrm{S3S4} \right) = \log \left( \frac{ \mathrm{I} \left( \left[ \mathrm{\ion{S}{iii}} \right] _{18\mu m} \right)  + \mathrm{I} \left( \left[ \mathrm{\ion{S}{iii}} \right] _{33\mu m} \right) }{  \mathrm{I} \left( \left[ \mathrm{\ion{S}{iv}} \right] _{10\mu m} \right) }  \right).
\end{equation}

Following the prescription by \citet{Perez-Diaz_2022}, the emission line [\ion{S}{iii}]{\ensuremath{\lambda}}33{\ensuremath{\mu}}m is only used for SFGs. The performance of these estimators is shown in Fig. \ref{s_estimators}. It is relevant to emphasize the importance of IR emission lines in estimating sulfur abundances. As shown in Fig. \ref{s_estimators_so}, S/O has little dependence on the behavior of the estimators, as intensity of IR emission lines mainly depends on the ionic abundance. However, this is no longer the case for optical lines, as they also depends on temperature, which translates in a dependence on the metallicity. Hence, estimators for sulfur based on optical emission lines (e.g. S23 Eq. \ref{S23}) are much more affected by the S/O assumed for the models.
\end{itemize}

\begin{figure*}
	\centering
	\includegraphics[width=0.7\hsize]{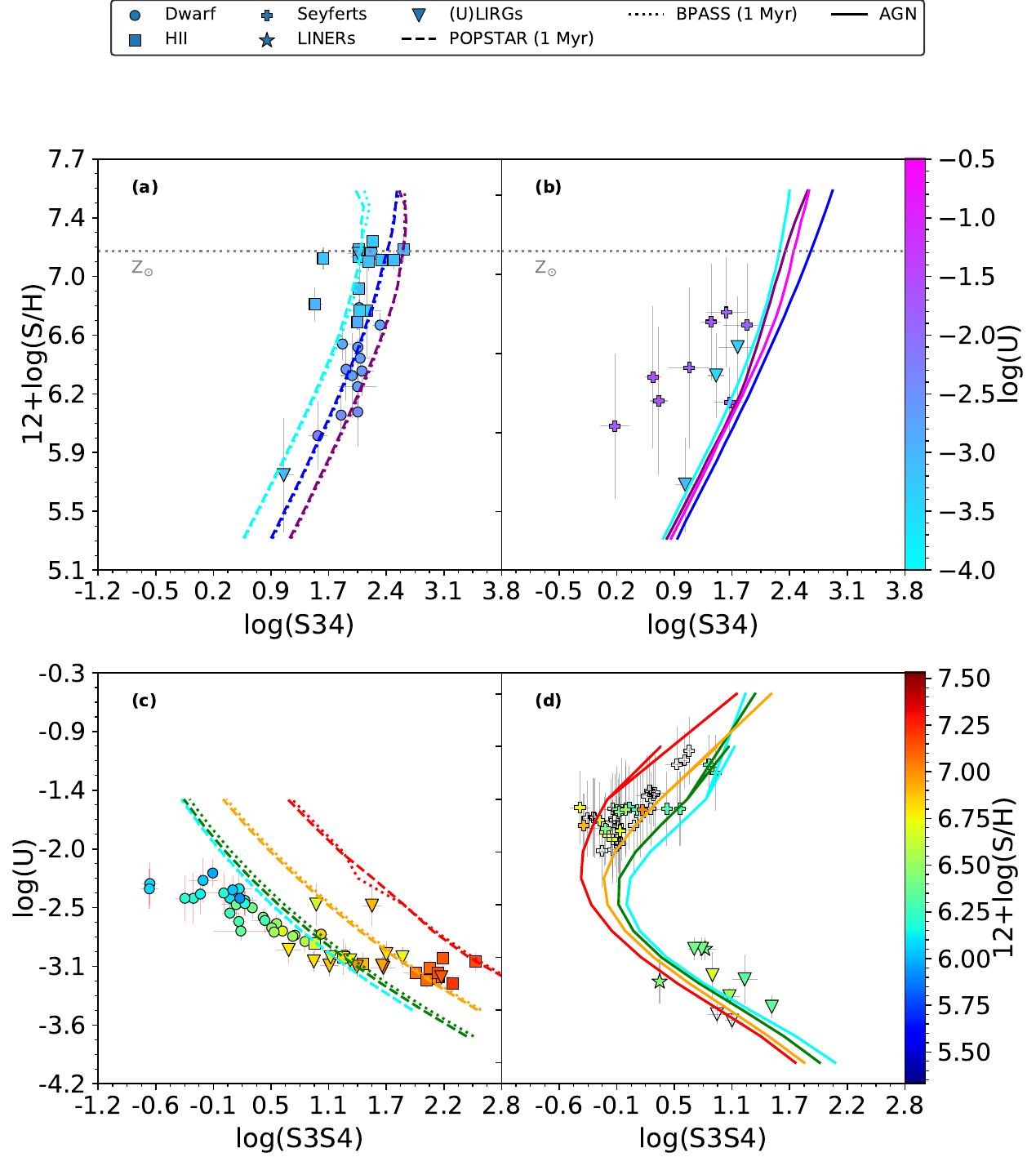}
	\caption{Performance of estimators based on sulfur emission lines. Panels (a) and (c) show the performance of S34 for 12+log(S/H) and S3S4 for log(U), respectively, for the case of SFGs. Panels (b) and (d) show the performance of the same estimators for AGNs. Models are shown following the same notation as in Fig.\, \ref{ar_estimators}.}
	\label{s_estimators}
\end{figure*}

\begin{figure*}
	\centering
	\includegraphics[width=0.7\hsize]{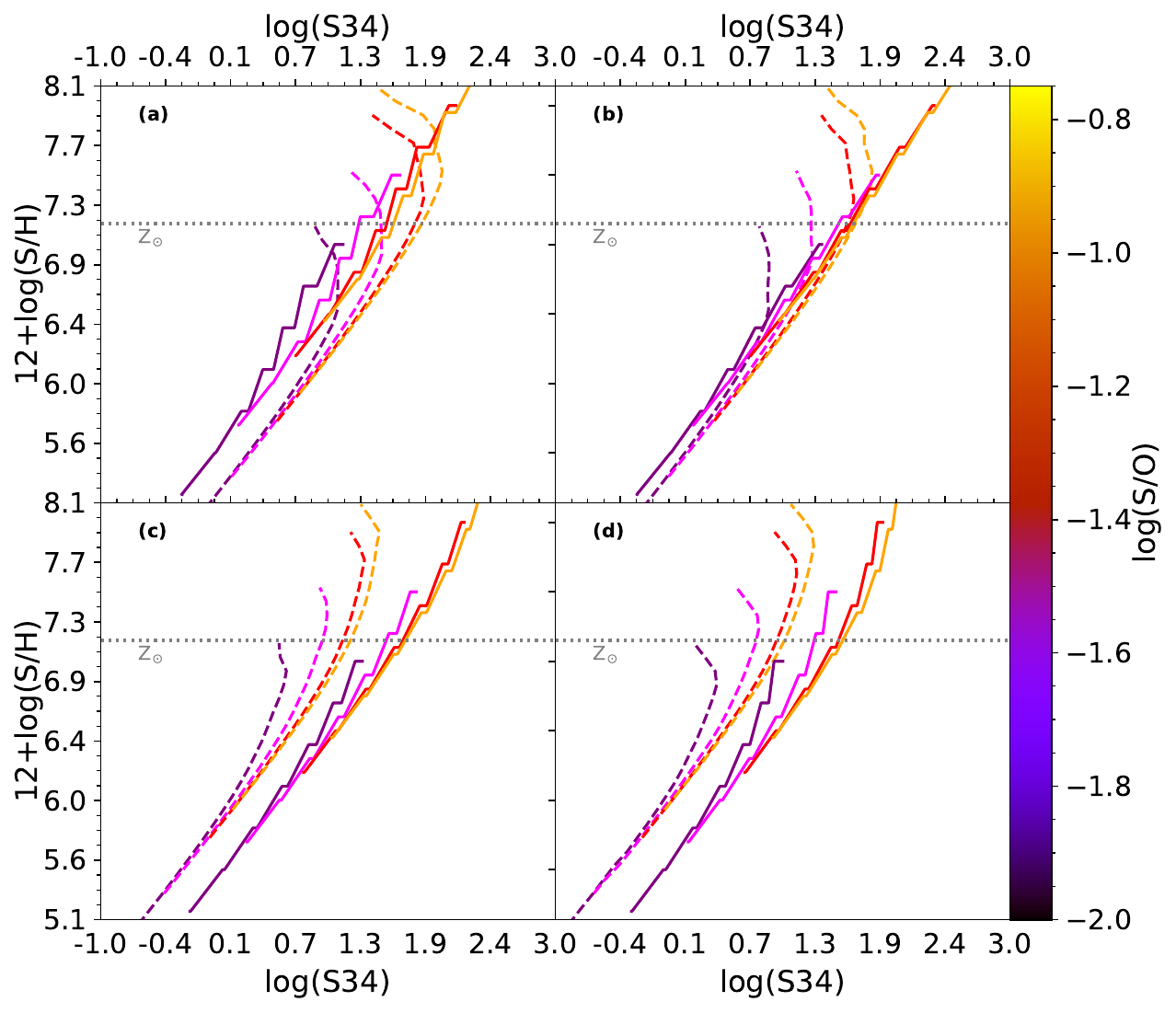}
	\caption{Performance of estimator S34 based on sulfur emission lines for models with S/O not fixed. Panel (a) shows models for a fixed value of the ionization parameter log(U)=-1.5, panel (b) shows models for log(U)=-2.0, panel (c) models with log(U)=-3.0 and panel (d) for log(U)=-3.5. Models are shown following the same notation as in Fig.\, \ref{ar_estimators}.}
	\label{s_estimators_so}
\end{figure*}

\begin{figure*}
	\centering
	\includegraphics[width=0.7\hsize]{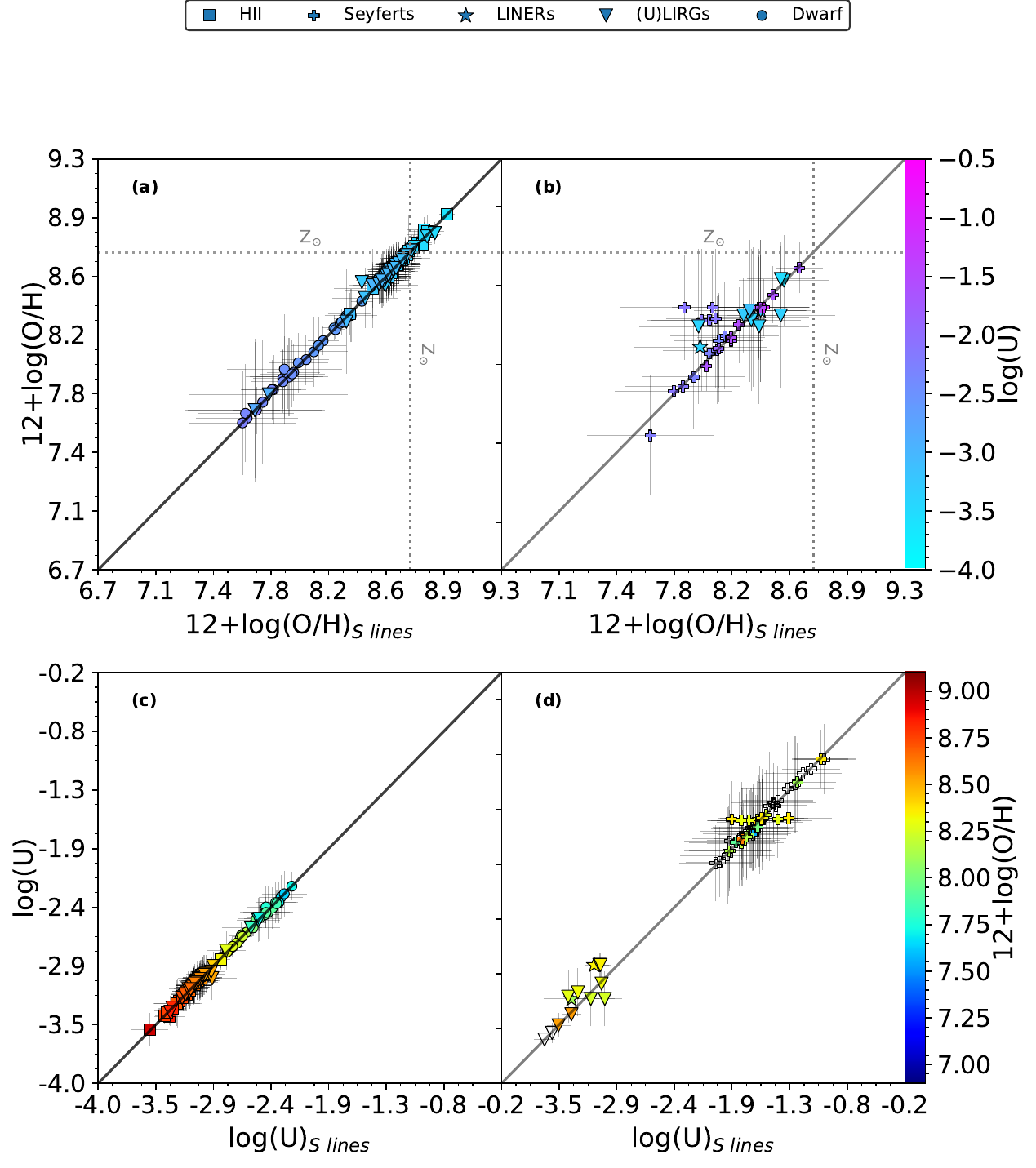}
	\caption{Comparison between the outputs from \textsc{HCm} v3.01 (x-axis) and \textsc{HCm} v3.1 (y-axis). Panels (a) and (b) show the comparison for 12+log(O/H) for SFGs and AGNs, respectively. Panels (c) and (d) show the comparison for log(U) for SFGs and AGNs, respectively.}
	\label{version_code}
\end{figure*}

As shown in Fig. \ref{version_code}, these new improvements in the code do not change significantly the results obtained from previous versions, but we obtain more information as now 12+log(S/H) is independently estimated. We notice that some AGNs seem to present slightly higher abundances when sulfur emission lines are no longer used in the oxygen estimation, which implies that S lines favor lower abundances (see Sec. \ref{sec3}).

\section{Results}
\label{sec3}
In this section we present the analysis of the chemical abundances obtained after applying \textsc{HCm-IR} to the selected samples presented in Sec. \ref{sec2}. The statistics of these results are shown in Table\,\ref{ir_results}.

\begin{table*}
	\caption{Statistics of the chemical abundances and log($U$) values derived from \textsc{HCm-IR} for our sample of galaxies.}
	\label{ir_results}
	\centering
	\begin{tabular}{ll|lll|lll|lll|lll}
		\multicolumn{2}{l}{} &
		\multicolumn{3}{|l}{\boldmath$12+\log_{10} \left( O/H \right) $} & \multicolumn{3}{|l}{\boldmath$12+\log_{10} \left( S/H \right) $} & \multicolumn{3}{|l}{\boldmath$\log_{10} \left( N/O \right) $} & \multicolumn{3}{|l}{\boldmath$\log_{10} \left( U \right) $}  \\ \hline \textbf{Sample} & \textbf{N\boldmath$_{tot}^{\circ}$} & \textbf{N\boldmath$^{\circ}$} & \textbf{Med.} & \textbf{Std. Dev.} & \textbf{N\boldmath$^{\circ}$} & \textbf{Med.} & \textbf{Std. Dev.} & \textbf{N\boldmath$^{\circ}$} & \textbf{Med.} & \textbf{Std. Dev.} & \textbf{N\boldmath$^{\circ}$} & \textbf{Med.} & \textbf{Std. Dev.} \\ \hline
		All SFGs & 131 & 128 & 8.57 & 0.30 & 59 & 6.68 & 0.38 & 22 & -0.93 & 0.20 & 71 & -2.97 & 0.34\\
		Dwarfs & 30 & 30 & 7.99 & 0.23 & 30 & 6.28 & 0.23 & 5 & -1.14 & 0.19 & 30 & -2.51 & 0.19 \\
		HII regions & 22 & 21 & 8.69 & 0.14 & 14 & 7.09 & 0.17 & 7 & -0.93 & 0.12 & 21 & -3.20 & 0.16 \\
		(U)LIRGs & 79 & 77 & 8.59 & 0.18 & 15 & 6.80 & 0.31 & 10 & -0.86 & 0.09 & 20 & -3.04 & 0.23 \\ \hline
		All AGNs & 73 & 36 & 8.28 & 0.21 & 36 & 6.41 & 0.38 & 35 & -0.83 & 0.16 & 65 & -1.68 & 0.68\\
		Seyferts & 58 & 25 & 8.25 & 0.22 & 25 & 6.57 & 0.37 & 22 & -0.79 & 0.19 & 52 & -1.60 & 0.21\\
		LINERs & 3 & 2 & 8.22 & 0.12 & 2 & 6.42 & 0.03 & 1 & -0.83 & 0.00 & 2 & -3.08 & 0.15\\
		(U)LIRGs & 12 & 9 & 8.31 & 0.11 & 9 & 6.36 & 0.36 & 12 & -0.88 & 0.07 & 11 & -3.23 & 0.22\\
	\end{tabular}
\end{table*}

\subsection{Nitrogen-to-oxygen abundance ratios}
\textsc{HCm-IR} performs a first independent iteration to estimate log(N/O). Since both the IR observables (i.e. N3O3 {\ensuremath \equiv} [\ion{N}{iii}]57\ensuremath{\mu}m/[\ion{O}{iii}]52\ensuremath{\mu}m ) and the procedure followed to determine N/O remain unchanged in this version, no significant difference is found in the log(N/O) distribution derived for the SFG and AGN samples when compared to studies based on previous versions of the code. In the case of SFGs, the median value of log(N/O) $\sim$ -0.9 is close to the solar ratio, in agreement with \citet{Fernandez-Ontiveros_2021}. Regarding AGNs, the distribution is almost identical to that reported by \citet{Perez-Diaz_2022}. Moreover, as IDEOS measurements do not cover the far-IR emission lines, essential to estimate log(N/O) \citep{Peng_2021, Fernandez-Ontiveros_2021, Perez-Diaz_2022}, the number of N/O measurements has not increased with respect to previous works.

\begin{figure*}
	\centering
	\includegraphics[width=0.8\hsize]{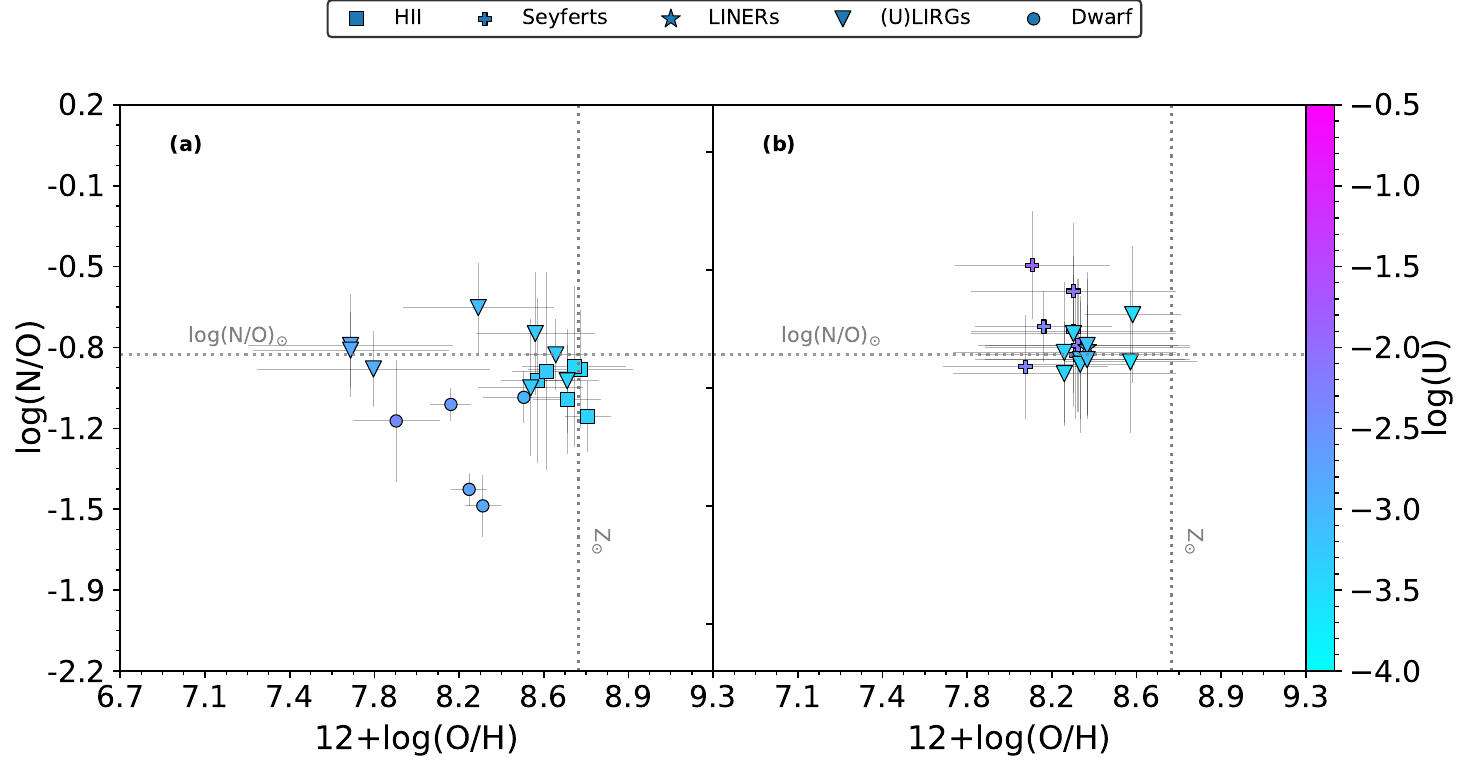}
	\caption{N/O vs O/H as derived from infrared emission lines for SFGs (a) and AGNs (b). The log($U$) values are given by the colorbar.}
	\label{no_oh}
\end{figure*}

Fig.\,\ref{no_oh} shows that the distribution of the IR-based log(N/O) vs. 12+log(O/H) values obtained for our sample deviates from the abundances obtained for local SFGs using optical lines \citep{Perez-Montero_2009, Andrews_2013, Perez-Montero_2014}. Our IR-based abundances cluster around the solar N/O ratio and do not show a trend with O/H. While this results seems to contradict previous studies \citep[e.g.][]{Spinoglio_2022}, we must bear in mind the low statistics of AGNs with N/O estimations. It is also important to note that while SFGs present values for N/O that spread from sub- to over-solar ratios, AGNs present either solar or slightly oversolar N/O ratios.

This result was also reported for the abundances of SFGs as derived from their IR lines \citep{Fernandez-Ontiveros_2021} and for AGNs \citep{Perez-Diaz_2022}, in both cases with calculations based on \textsc{HCm-IR}. Since 12+log(O/H) is now obtained without considering the IR sulfur lines, in contrast to previous studies, we conclude that these behaviors on the N/O-O/H diagram persist even when the information is solely obtained from O, Ne and Ar emission lines.

\subsection{Sulfur and oxygen abundances}

The analysis of the total oxygen abundance in our sample of SFGs is in overall agreement with \citet{Fernandez-Ontiveros_2021}. We found the lowest 12+log(O/H) average values in dwarf galaxies (12+log(O/H)$\sim$8.0), with sub-solar abundances for (U)LIRGs (12+log(O/H)$\sim$8.5) and the highest values for HII regions (12+log(O/H)$\sim$8.7). Nevertheless, our larger (ULIRG) sample presents a slightly higher median oxygen abundance when compared to the smaller sample of 12 objects in \citet{Fernandez-Ontiveros_2021}. The agreement with previous determinations imply that the independent estimation of S in the last version of \textsc{HCm-IR} does not significantly change the abundances of O and N, as in the bayesian-like procedure sulfur estimators are weighted among many others, reducing any possible bias. Regarding the derived average sulfur content 12+log(S/H) in our sample of SFGs, we obtained a similar behavior as that for oxygen, being the lowest value found in dwarfs (12+log(S/H)$\sim$6.3), followed by (U)LIRGs (12+log(S/H)$\sim$6.8) and HII regions (12+log(S/H)$\sim$7.1).

The 12+log(O/H) values obtained for AGNs are also consistent with previous studies \citep{Perez-Diaz_2022}, as there is not a significant increase of the oxygen content for high-ionisation (Seyferts) AGNs. This is supported by the larger statistics after including the AGNs from IDEOS. Hence, we can conclude that in terms of chemical content, both samples from IDEOS and from \citet{Perez-Diaz_2022}, are similar. Overall, the whole AGN sample presents similar median values for both 12+log(O/H)\ensuremath{\sim}8.25 and 12+log(S/H)\ensuremath{\sim}6.5 for all considered subtypes (Seyferts, LINERs and (U)LIRGs).

\begin{figure*}
	\centering
	\includegraphics[width=0.8\hsize]{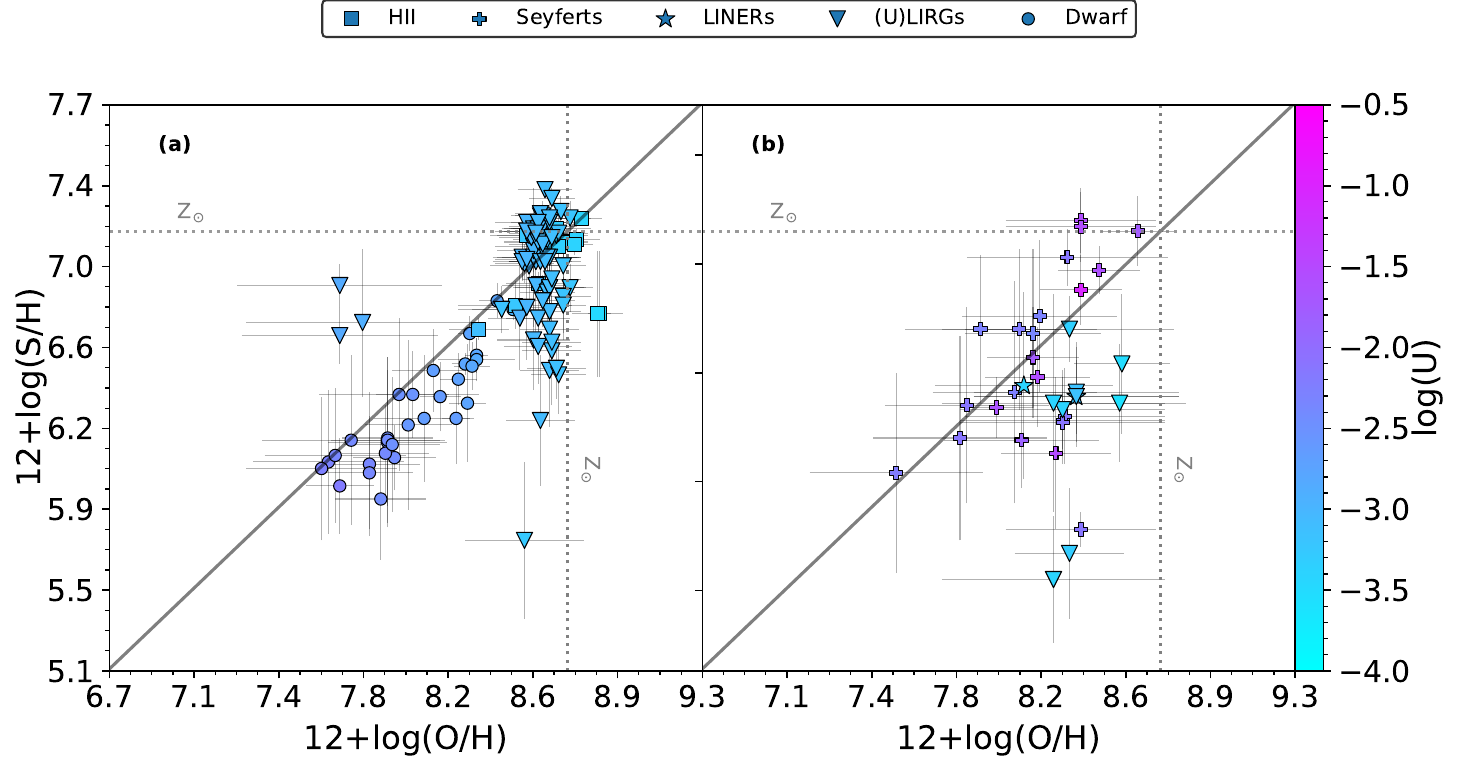}
	\caption{S/H vs O/H as derived from infrared emission lines for SFGs (a) and AGNs (b). The log($U$) values are given by the colorbar. The solid line represents the solar proportion. We notice in panel (a) that (U)LIRGs with high S/H values near the lowest values for O/H are the same galaxies that are experiencing the {\it deep-diving} phase as reported by \citet{Perez-Diaz_2023}.}
	\label{sh_oh}
\end{figure*}

Finally, we also explored the relation between the ionisation parameter log(U) and the derived chemical abundances. Theoretically, if massive stars are the sources of ionisation, an anticorrelation between metallicity and ionisation is expected as a consequence of: 1) stars become cooler as a result of wind and enhanced line blanketing \citep{Massey_2005}; and, 2) an increase in the stellar atmosphere content leads to higher photon scattering, which later translates to a more efficient conversion of the luminosity energy
into the mechanical energy in winds \citep{Dopita_2006}. As presented in Fig. \ref{U_vs_Z}, we did obtain an anticorrelation for SFGs which is stronger for oxygen (pearson coefficient correlation is r\ensuremath{\sim}-0.98) than for sulfur (pearson coefficient correlation is r\ensuremath{\sim} -0.8). When analysing AGNs, we do not obtain any relation between metallicity and ionisation parameter, in agreement with previous studies \citep[e.g.][]{Perez-Montero_2019, Perez-Diaz_2021, Perez-Diaz_2022}. Moreover, the lack of correlation between U and O/H (and S/H) is obtained in galaxies characterised by high ionisation parameters (log(U) > -2.5 such as Seyferts) but also in AGN-dominated (U)LIRGs with low ionisation parameters (log(U) < -2.5), as shown in Fig. \ref{sh_oh}.

\subsection{Sulfur-to-oxygen abundance ratios}

The median log(S/O) values obtained for our samples of SFGs (-1.89) and AGNs (-1.87) are lower than the solar ratio log(S/O)\ensuremath{_{\odot}} = -1.57 \citep{Asplund_2009}. From Fig. \ref{sh_oh} we conclude that, although median values deviate by a factor of 0.3 dex from the solar ratio, there are found higher offsets in many of the (U)LIRGs (of our sample, dominated either by star-forming or AGN activity), as their sulfur abundances are significantly lower than the expected value for their corresponding oxygen estimations.

\begin{figure*}
	\centering
	\includegraphics[width=0.8\hsize]{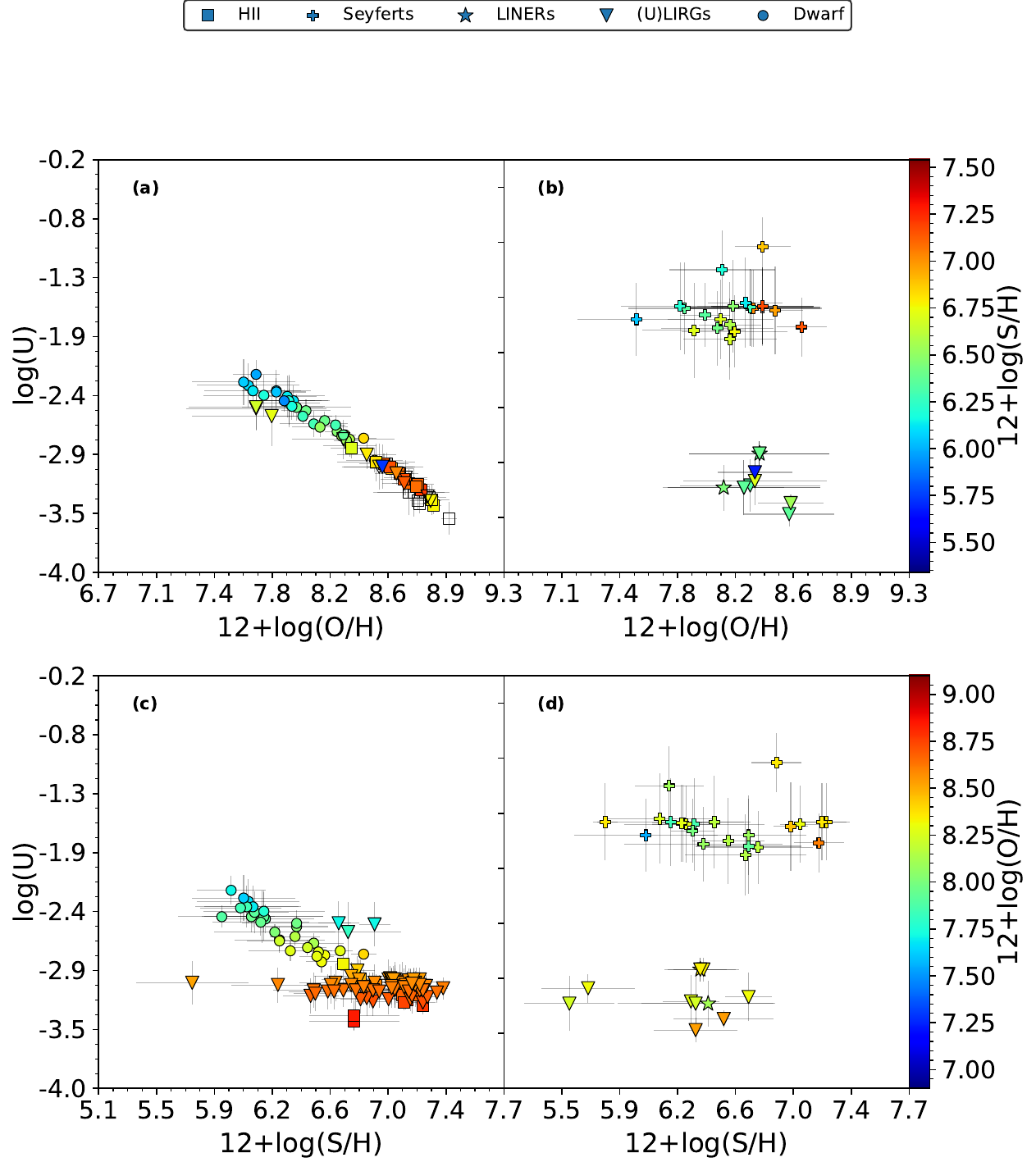}
	\caption{Variation of the ionisation parameter log(U) as a function of the chemical abundance ratios 12+log(O/H), panels (a) and (b), and 12+log(S/H), panels (c) and (d). Panels (a) and (c) show results from SFGs, while panels (b) and (d) present AGNs.}
	\label{U_vs_Z}
\end{figure*}

\begin{figure}
	\centering
	\includegraphics[width=0.8\hsize]{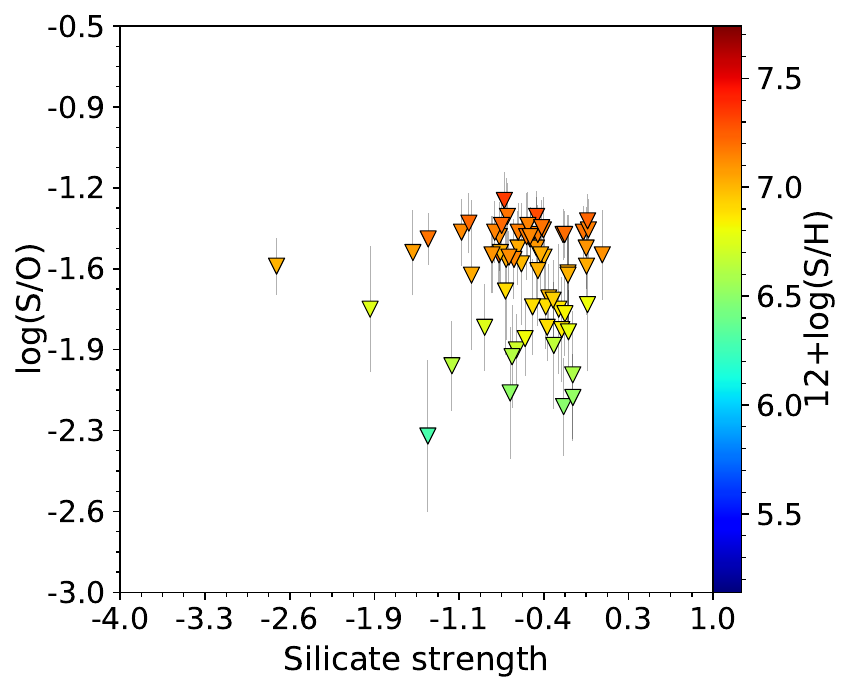}
	\caption{S/O vs the silicate strength in the 10-12.58\ensuremath{\mu}m range as measured in IDEOS \citep{Spoon_2022}. The colorbar shows the values for the 12+log(S/H) chemical abundance.}
	\label{silicate}
\end{figure}

First of all, we explored the possibility that these deviations in the log(S/O) chemical abundance ratio could be  caused by uncertainties in the measurement of the [S{\sc iv}]\ensuremath{\lambda}10\ensuremath{\mu}m emission line.  
Among these, the flux of this line could be affected as a consequence of the presence of a silicate feature detected in the 10-12.58\ensuremath{\mu}m range \citep{Spoon_2022}. From Fig. \ref{silicate} we can conclude that the silicate strength does not play any role in the estimated values for log(S/O). However, we must bear in mind that in galaxies dominated by AGN activity the silicate strength provides information on the extinction that is affecting the continuum emission, whereas the line emission can come from more extended parts.

To further explore the above reported deviations in the chemical abundance ratio log(S/O) for our sample, we studied its behavior as a function of the oxygen abundance. As shown in Fig. \ref{so_oh}, the deviations in log(S/O) are found in either the low-metallicity regime (12+log(O/H) $<$ 8.0), as is the case of SFGs, or in the high-metallicity regime (12+log(O/H) $>$ 8.4), for both SFGs and AGNs. In the two metallicity regimes, (U)LIRGs are the galaxies driving these high deviations.

\begin{figure*}
	\centering
	\includegraphics[width=0.8\hsize]{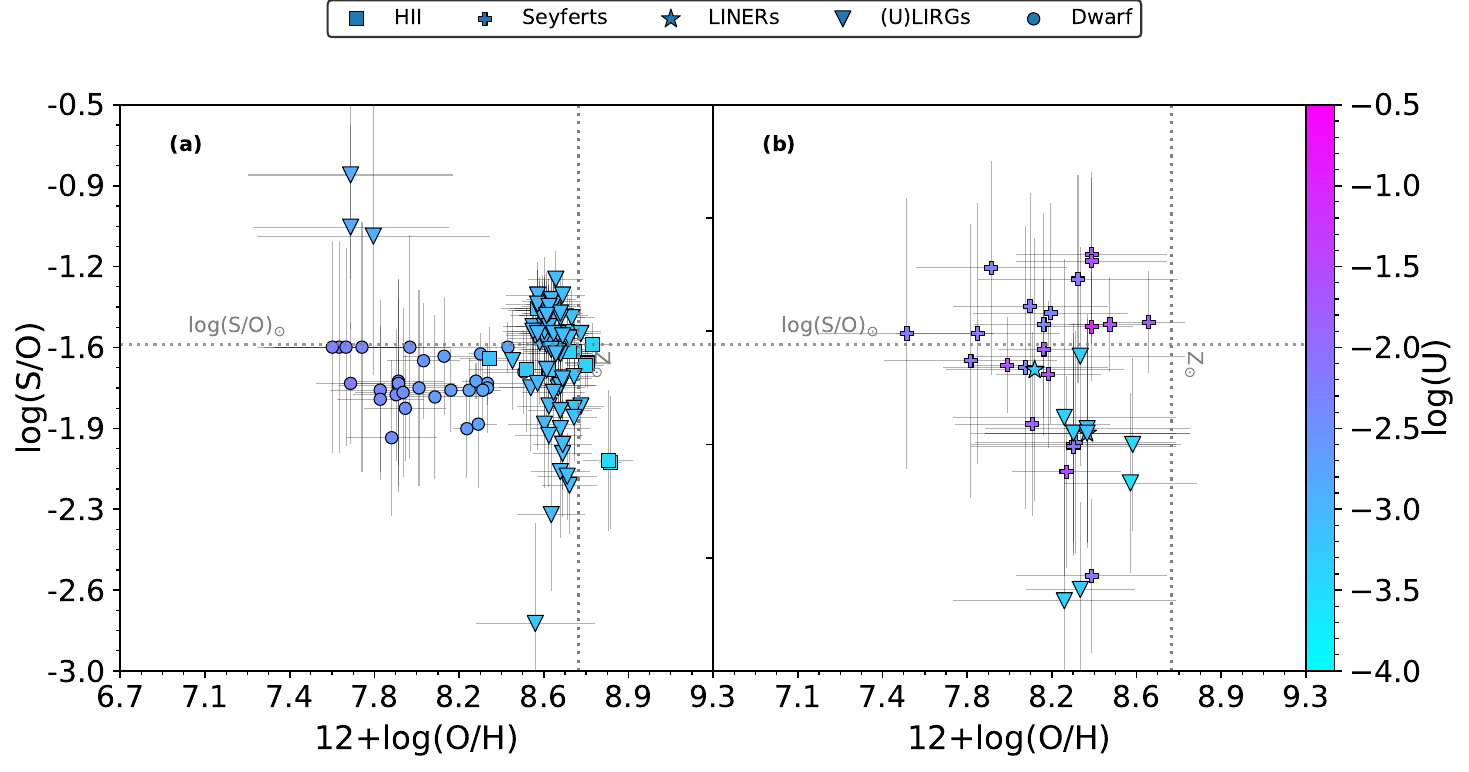}
	\caption{S/O vs O/H as derived from the infrared emission lines for SFGs (a) and AGNs (b). The log($U$) values are given by the colorbar. Again, (U)LIRGs that show the highest S/O ratios are the same galaxies that undergoing their {\it deep-diving} phase as reported by \citet{Perez-Diaz_2023}.}
	\label{so_oh}
\end{figure*}

In the specific case of SFGs, we explored if these deviations correlate with other physical quantities such as stellar mass (M\ensuremath{_{*}}) or star-formation rate (SFR), as galaxy mass assembly is known to play an important role in the chemical enrichment of galaxies \citep{Curti_2017, Maiolino_2019}, materialised in the well-known Mass-Metallicity Relation \citep[MZR,][]{Lequeux_1979, Tremonti_2004, Andrews_2013}. Although this relation is regulated by many processes such as galaxy environment \citep{Peng_2014}, secular evolution \citep{Somerville_2015}, star-formation and AGN feedback \citep{Blanc_2019, Thomas_2019} or stellar age \citep{Duarte-Puertas_2022}, we only analysed those quantities which are explicitly involved in this connection between chemical enrichment and galaxy mass assembly: M\ensuremath{_{*}}\citep{Lequeux_1979, Tremonti_2004} and SFR \citep[SFR,][]{Mannucci_2010, Curti_2017}.

Fig. \ref{so_mass} shows that deviations from the solar ratio log(S/O)$_{\odot}\sim$ -1.57 are observed mostly in those galaxies with high M\ensuremath{_{*}} ($>$10\ensuremath{^{11}}\,M\ensuremath{_{\odot}}) and high SFR (>90 M\ensuremath{_{\odot }}\,yr\ensuremath{^{-1}}).

\begin{figure*}
	\centering
	\includegraphics[width=0.8\hsize]{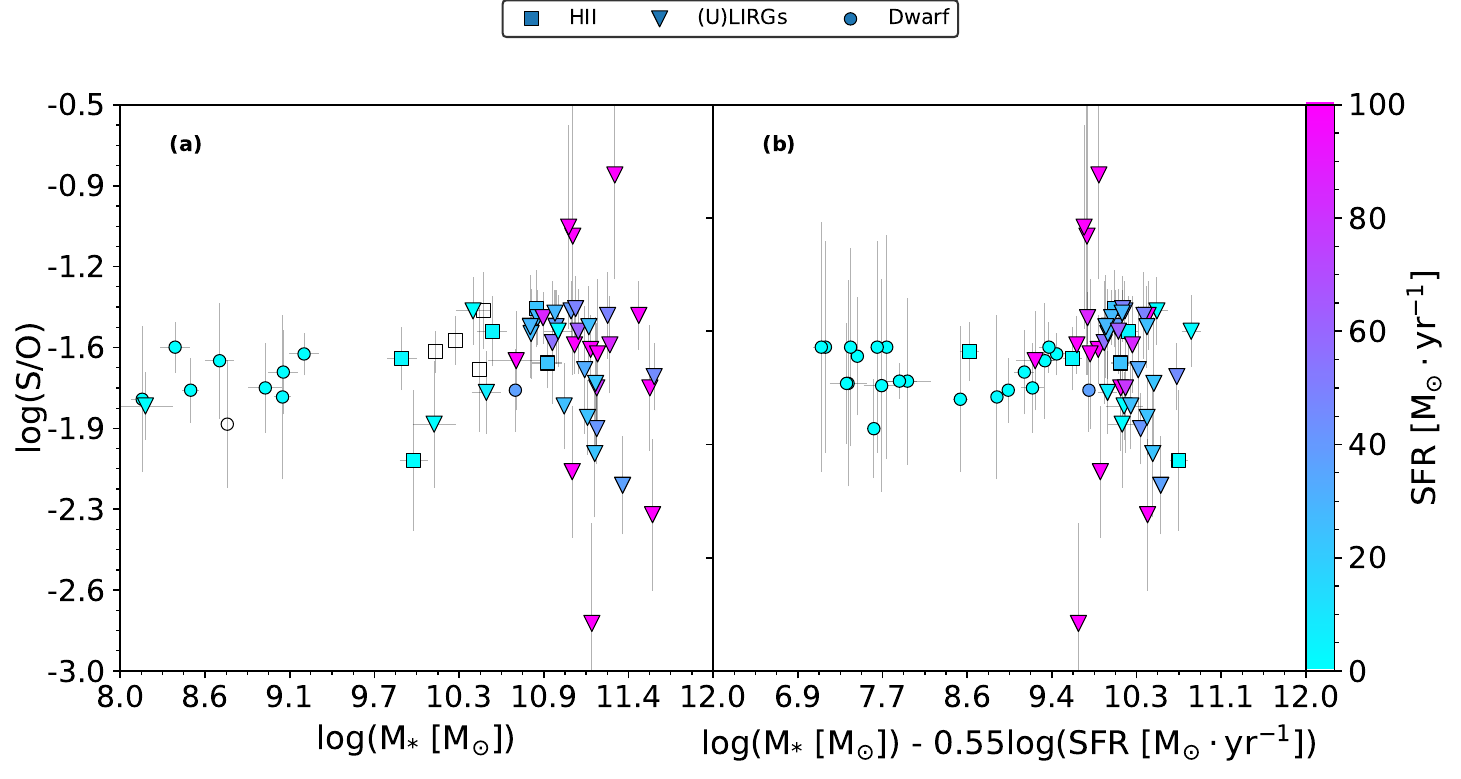}
	\caption{(a) S/O vs M\ensuremath{_{*}}. (b) S/O vs M\ensuremath{_{*}} corrected from SFR as proposed by \citet{Curti_2017}. The SFR values are given by the colorbar.}
	\label{so_mass}
\end{figure*}

\section{Discussion}
\label{sec4}
\subsection{Information from sulfur emission lines}

Unlike previous analyses of chemical abundances of the gas-phase using IR lines, which assume of a constant and universal S/O ratio, our work avoids S emission lines --\,namely [\ion{S}{iv}] {\ensuremath{\lambda}}10{\ensuremath{\mu}}m, [\ion{S}{iii}]{\ensuremath{\lambda}}18{\ensuremath{\mu}}m and [\ion{S}{iii}]{\ensuremath{\lambda}}33{\ensuremath{\mu}}m\,-- to estimate 12+log(O/H). Instead, the sulfur lines are used in this work to estimate 12+log(S/H) independently. Given the quantity and variety of other emission lines observed in the IR range (Ne, Ar, O), sulfur emission lines do not play a critical role for the estimation of the oxygen abundance, as evidenced by the high consistency between our results and previous studies using all lines (see Fig. \ref{version_code}).

While IR emission lines are extremely useful due to the fact that they are much less affected by temperature or dust extinction, their dependence on density should be taken into account. Far-IR emission lines present low critical densities \citep[e.g.][]{Perez-Diaz_2022}, which implies that their fluxes can be affected even for the average conditions of the ISM in both SFGs (n{\ensuremath{_{e}\sim}}100\,cm$^{-3}$) and AGNs (n{\ensuremath{_{e}\sim}}500\,cm$^{-3}$). This is not the case for IR sulfur emission lines, as their critical densities (n\ensuremath{_{c}}([\ion{s}{iv}]\ensuremath{_{10\mu m}})\ensuremath{\sim}5.6\ensuremath{\cdot 10^{4}}\,cm$^{-3}$, n\ensuremath{_{c}}([\ion{s}{iii}]\ensuremath{_{18\mu m}})\ensuremath{\sim}1.2\ensuremath{\cdot 10^{4}}\,cm$^{-3}$ and n\ensuremath{_{c}}([\ion{s}{iii}]\ensuremath{_{33\mu m}})\ensuremath{\sim}1417\,cm$^{-3}$; \citealt{Perez-Diaz_2022}) are well above such conditions of the ISM.

Furthermore, as the derivation of chemical abundances based on the IR range also involves intermediate and highly ionised species (S$^{2+}$, S$^{3+}$), the influence from diffuse ionised gas (DIG; \citealt{Reynolds_1985, Domgorgen_Mathis_1994, Galarza_1999, Zurita_2000, Haffner_2009}), whose contribution is only relevant for low-excitation lines, is expected to be negligible in the abundance calculation. In addition, the detection of high-excitation IR lines is also extremely important in the case of AGNs, especially in high-luminosity AGNs such as Seyferts or quasars, since highly ionised species are needed to correctly trace the chemical content of the gas-phase ISM, and these ions produce emission lines which are easily detected in the IR range.

\subsection{Deviations from the S/O solar ratio}
Our analysis of the chemical abundance ratio log(S/O) in the selected sample, calculated from the independent estimation of sulfur and oxygen abundances, reveals that many galaxies present values close to the solar ratio. As a matter of fact, about 70\% of our sample of SFGs are within 0.2\,dex of the solar ratio. However, this value drops to 44\% when AGNs are considered. Focusing on the galaxies whose S/O ratio clearly deviates from the solar proportion (> 0.3\,dex), we found that 53\% of the AGNs and 17\% of the SFGs present such large deviations.

Regarding the different subtypes of galaxies, (U)LIRGs present the highest deviations (see Fig. \ref{sh_oh}). Specifically, we found that (U)LIRGs with an oxygen content close to the solar value (12+log(O/H)\ensuremath{_{\odot}\sim}8.69), show a spread in S/O ratios of more than an order of magnitude (from -1.2 to -2.3), suggesting a strong variation of the sulfur content for galaxies with similar stellar masses M\ensuremath{_{*} \sim }10\ensuremath{^{11}}M\ensuremath{_{\odot }} and oxygen abundances 12+log(O/H)\ensuremath{\sim}8.6 (see Fig. \ref{so_mass}).

Among the (U)LIRGs that deviate from the solar log(S/O) ratio are those with very low abundances (12+log(O/H) < 8.2), as compared to the rest of galaxies of the same type. These (U)LIRGs possibly undergo a large excursion or {\it deep-dive} beneath the mass-metallicity relation (MZR) due to massive infalls of metal-poor gas \citep{Perez-Diaz_2023}. According to our results, sulfur abundances are higher in these objects, leading to over-solar log(S/O)$>$-1.2 ratios. This behavior is similar to that reported by several authors using optical lines for different SFG samples \citep{Diaz_1991, Pilyugin_2006, Diaz_2022}.

Diverse mechanisms able to drive the observed log(S/O) deviation in SFGs have been proposed, including: 1) changes in the initial mass function (IMF) enhancing the formation of stars with masses between 12\,M\ensuremath{_{\odot }} and 20\,M\ensuremath{_{\odot }}, which are the major producers of S via burning of O and Si \citep{Diaz_2022}; and, 2) metal enrichment from Type Ia Supernovae increasing the sulfur yield \citep{Iwamoto_1999}. The first scenario could be favored in the extreme star formation conditions that characterize {\it deep-diving} (U)LIRGs, as larger SFRs helps in increasing the number of stars with M\,\ensuremath{>}\,10\,M\ensuremath{_{\odot}}. Moreover, if (U)LIRGs might have experienced a change in their IMF, which has not been reported yet, then this effect could be amplified. On the other hand, the second mechanism proposed could be important in galaxies with strong outflows, for instance \citet{Perez-Diaz_2023} discuss how strong feedback from the extreme episode of star formation is required for (U)LIRGs to end their {\it deep-diving} phase. 

However, as shown in Fig. \ref{so_mass}, these deviations found in massive galaxies are not always associated to an increase of S, as we also observed extremely low log(S/O) abundances for similar conditions. Moreover, these scenarios do not explain why there is such a strong variation on the sulfur content for galaxies with similar oxygen abundances and stellar properties such as mass or SFR. Another possible mechanism that could explain these lower abundances of S/H might be related to the formation of ice and dust grains that capture S in the most dense cores of molecular clouds \citep{Hily-Blant_2022}. Indeed, recent studies show that sulfur abundances in these cold dense clouds are much lower than those reported in the ionised ISM \citep[e.g.][]{Fuente_2023}, and these cold regions can be traced by IR emission lines while they remain unobserved from optical spectroscopy. Nevertheless, we cannot explore this scenario with the current IR data.

AGNs show an analogous behavior with S/O abundances spread over a wide range for galaxies characterised by similar oxygen abundances. This was also reported by \citet{Dors_2023} using optical emission lines, however the information from S$^{3+}$ was missing, which plays an important role to derive the total S abundance in the case of strong AGNs such as Seyferts. As in the case of SFGs, it is unclear whether the proposed scenarios of chemical enrichment from both stellar nucleosynthesis and feedback might explain the observed deviations. Nevertheless, we report the absence of AGNs with clear over-solar log(S/O) ratios, unlike {\it deep-diving} (U)LIRGs, implying either that galaxies classified as AGNs in our sample cannot meet the conditions to present such values, or that there is an observational bias towards objects with solar-to-subsolar log(S/O) ratios.

In the near future, the analysis of spatially-resolved spectroscopic observations in the IR range for both SFGs and AGNs will shed light to better disentangle if the mechanisms and their extent proposed to explain the observed S/O variation are the same in both type of objects.

\subsection{The role of electron density}
The emissivities of the IR emission lines are significantly less dependent on temperature when compared with the optical transitions \citep[e.g.][]{Fernandez-Ontiveros_2021}. However, the former are more affected by the electron density (n\ensuremath{_{e}}) due to the lower critical densities (n\ensuremath{_{c}}) of the IR transitions --\,i.e. the density at which collisional de-excitations equal radiative transitions. For instance, [\ion{O}{iii}]88\ensuremath{\mu}m has n\ensuremath{_{c}} = 501 cm\ensuremath{^{-3}}, whereas [\ion{O}{iii}]5007\ensuremath{\AA} has 6.9\ensuremath{\times 10^{5}} cm\ensuremath{^{-3}}. As shown in \citet{Perez-Diaz_2022}, most of the IR emission lines used in this work have n$_{c}$ well above the expected densities in SFGs ($\sim100$ cm$^{-3}$) and AGNs ($\sim500$ cm$^{-3}$), and therefore effects from the density conditions of the ISM are not critical for chemical abundance estimations. Additionally, \textsc{HCm-IR} \citep{Fernandez-Ontiveros_2021,Perez-Diaz_2022} takes into account this information in the calculations and uses only those IR transitions whose n\ensuremath{_{c}} are above the expected densities for the ionised ISM in each case. Nevertheless, to test the robustness of the S/O relative abundances obtained, we investigate in this section a possible dependency on the gas density.

For this purpose, we evaluate two ratios that are extremely sensitive to n\ensuremath{_{e}}: [\ion{S}{iii}]33\ensuremath{\mu}m/[\ion{S}{iii}]18\ensuremath{\mu}m and [\ion{Ne}{v}]24\ensuremath{\mu}m/[\ion{Ne}{v}]14\ensuremath{\mu}m. The former is used to trace densities in SFGs, because it is the only available set of emission lines from the same ionic species in the mid-IR. The latter is used to measure densities in AGNs, as this ratio involves a highly ionised ion, thus avoiding possible contamination from star formation activity. In Fig. \ref{so_ne} we present the log(S/O) values as a function of these two emission line ratios.

\begin{figure*}
	\centering
	\includegraphics[width=0.8\hsize]{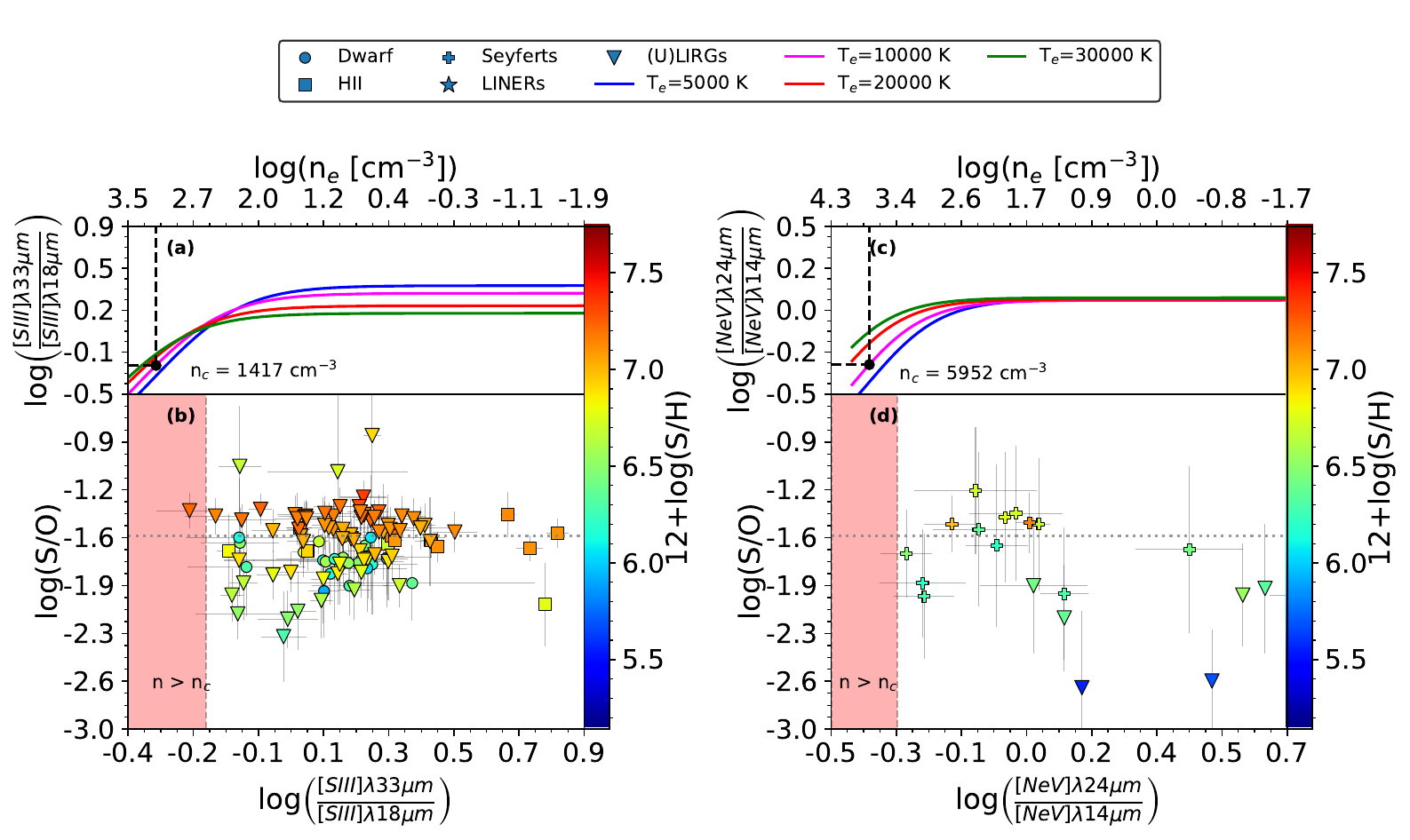}
	\caption{Dependence of S/O on the electron density. Panels (a) and (c) show the theoretical behavior of the emission line ratios [\ion{S}{iii}]33\ensuremath{\mu}m/[\ion{S}{iii}]18\ensuremath{\mu}m and [\ion{Ne}{v}]24\ensuremath{\mu}m/[\ion{Ne}{v}]14\ensuremath{\mu}m respectively as a function of density for different values of the electron temperature, as computed from \textsc{Pyneb} \citep{Luridiana_2015}. Panel (b) shows the behavior of log(S/O) as a function of [\ion{S}{iii}]33\ensuremath{\mu}m/[\ion{S}{iii}]18\ensuremath{\mu}m for SFGs. Panel (d) shows the behavior of log(S/O) as a function of [\ion{Ne}{v}]24\ensuremath{\mu}m/[\ion{Ne}{v}]14\ensuremath{\mu}m for AGNs. Colorbar shows the values of 12+log(S/H) for both samples. Critical densities were computed assuming T\ensuremath{_{e}}=10\,000 K and are those associated to the emission line with the lowest value.}
	\label{so_ne}
\end{figure*}

The lack of correlation between log(S/O) and n\ensuremath{_{e}} is shown in Fig. \ref{so_ne}, suggesting that density variations are not causing the deviations from the solar ratio. A few SFGs (five) and AGNs (three) present densities which are compatible within the errors with the critical density regime (n\ensuremath{_{e}} > n\ensuremath{_{c}}), although the density uncertainties in all cases are compatible with lower values. Analyzing the statistics for each sample, we found that SFGs and AGNs present median values of n\ensuremath{_{e}\sim}200 cm\ensuremath{^{-3}} and \ensuremath{\sim}1000 cm\ensuremath{^{-3}}, respectively, adopting an electron temperature of T\ensuremath{_{e}=10}\ensuremath{^{4}} K. Nevertheless, we note that given the uncertainties in the emission line ratios involved, these results are still compatible with the typical ISM densities for each case.

On the other hand, there are some considerations that may impact the determination of electron densities. For instance, it has been probed that planetary nebulae present density inhomogeneities across the gas-phase ISM \citep[e.g.][]{Seaton_1957, Flower_1969, Harrington_1969, Pequignot_1979, Rubin_1989}. Additionally, the action of shocks in the ISM might also induce variations in the density distribution \citep[e.g.][]{Dopita_1976, Dopita_1977, Dopita_1977b, Contini_1983, Aldrovandi_1985}. While these effects can play an important role in the determination of the electron temperature T\ensuremath{_{e}} for a direct estimation of chemical abundances using CELs, \textsc{HCm-IR} does not make any prior estimation on T\ensuremath{_{e}}. \citet{Perez-Montero_2019} show that a variation on the electron density by a factor of 4 has a negligible effect on the chemical abundances derived using \textsc{HCm}. In the case of IR estimators, such as S34, Ne23 or Ne235, only remarkable differences (higher than 0.3 dex) are found when the densities are changed from 100 cm\ensuremath{^{-3}} for SFGs and 500 cm\ensuremath{^{-3}} for AGNs to 10\,000 cm\ensuremath{^{-3}} in both cases, according to photoionization models computed with \textsc{Cloudy} v17 \citep{Ferland_2017}. This value is well above the critical density for most of the IR emission lines considered, and there are no galaxies in this regime in our sample of SFGs and AGNs, as shown in Fig. \ref{so_ne}. Therefore, we conclude that the density conditions have no significant impact on the chemical abundances of 12+log(S/H) and 12+log(O/H) estimated in this work.

Finally, we note that a few SFGs and AGNs show values above the expected ratio in the low-density regime (i.e., n\ensuremath{_{e} \rightarrow 0}). Nevertheless, the deviations from the log(S/O) solar ratio are also found in the density regime between 100 cm\ensuremath{^{-3}} and 1000 cm\ensuremath{^{-3}}, supporting the idea that these differences should not be driven by variations in the gas density, but are instead intrinsic to the chemical composition of the ISM.

\section{Conclusions}
\label{sec5}
In this work we performed, for the first time, a systematic analysis of the chemical abundances estimated from IR emission lines in a sample of galaxies including both SFGs and AGNs, providing an independent estimation of the oxygen, nitrogen and sulfur abundances for the widest sample of galaxies with IR spectroscopic observations combining {\it Spitzer}, {\it Herschel}, AKARI and SOFIA. When compared with previous studies of chemical abundances from IR emission lines, we found an agreement in the estimated oxygen and nitrogen abundances.

While most of the galaxies in the sample are characterised by a solar log(S/O) ratio, we report that galaxies from low to solar oxygen abundances present large deviations in log(S/O). In the first case, galaxies present higher sulfur abundances, which is consistent with studies based on optical emission lines. Among them, (U)LIRGs are characterised by high stellar masses, high star formation rates and some are reported to be experiencing an infall of metal-poor gas, strengthening the hypothesis that the large S/O ratios may be
driven by stellar nucleosynthesis. In the case of galaxies with solar-like abundances, we find that (U)LIRGs present S/O ratios that span over a very wide range, whereas their stellar properties remain similar, implying a possible additional sulfur production channel or a more complex picture on stellar nucleosynthesis. We also tested whether these deviations from the S/O solar ratio are driven by the density conditions of the ISM, and we conclude that density is not a driver of such deviations.

Our results for AGNs are similar to those observed in SFGs, although in this case we report more galaxies deviating from the log(S/O) solar ratio. We found that the (U)LIRGs dominated by AGN activity and some Seyferts present extremely low log(S/O) ratios, while they show almost solar oxygen abundances. Unlike the SFGs, we do not find any AGN with subsolar abundances and high log(S/O) ratios, although this could be due to the low number of low-metallicity AGNs in our sample.

On the view of these results, while the use of IR sulfur emission lines to constrain the overall metallicity of the gas-phase ISM is not required when transitions from other ionic species are detected (neon, argon, oxygen, ...), using only sulfur lines can lead to an underestimation of 12+log(O/H) in the solar (and over solar) regime, and an overestimation for galaxies which are actually characterised by a depressed oxygen content.

\begin{acknowledgements}
We acknowledge support from the Spanish MINECO grant PID2022-136598NB-C32. We also acknowledge  financial  support  from  the  State Agency for Research of the Spanish MCIU through the "Center of Excellence Severo Ochoa" award to the Instituto de Astrof\'{\i }sica de Andaluc\'{\i }a (SEV-2017-0709). J.A.F.O., A.H.C. and R.A. acknowledge financial support by the Spanish Ministry of Science and Innovation (MCIN/AEI/10.13039/501100011033) and ``ERDF A way of making Europe'' though the grant PID2021-124918NB-C44; MCIN and the European Union -- NextGenerationEU through the Recovery and Resilience Facility project ICTS-MRR-2021-03-CEFCA. We acknowledge the fruitful discussions with our research team. We thank the anonymous referee for the constructive report that improved this manuscript. E.P.M. acknowledges the assistance from his guide dog, Rocko, without whose daily help this work would have been much more difficult.
\end{acknowledgements}

\bibliographystyle{bibtex/aa} 
\bibliography{hcm}

\begin{appendix}
\section{Data}
\label{sec6}
We present in this appendix the full dataset of mid- to far-IR spectroscopy of our samples of SFGs (Table \ref{TabA1}) and AGNs (Table \ref{TabA2}). Table \ref{TabA3} and Table \ref{TabA4} show our estimations from IR emission lines (for SFGs and AGNs respectively) of chemical abundances and ionization parameters for our sample.

\begin{landscape}
	\begin{table}
		\caption{List of IR fluxes and stellar properties for our sample of SFGs.}\label{TabA1}
		\centering
		\begin{tabular}{llllllllllll}
			\textbf{Name} & \textbf{RA} & \textbf{De} & \textbf{z} & \textbf{Type} & \boldmath$\mathrm{Br}_{\alpha }$ & \textbf{[\ion{Ar}{ii}]7$\mu$ m} & ... & \boldmath$\log \left( \mathrm{M}_{e} \left[ \mathrm{M}_{\odot} \right] \right)$ & \boldmath$\mathrm{SFR}\left[\mathrm{M}_{\odot}\cdot\mathrm{yr}^{-1}\right]$ &  \textbf{Ref.} & \textbf{Ref. ste.} \\  
			\textbf{(1)} & \textbf{(2)} & \textbf{(3)} & \textbf{(4)} & \textbf{(5)} & \textbf{(6)} & \textbf{(7)} & ... & \textbf{(20)}  & \textbf{(21)} & \textbf{(22)} & \textbf{(23)} \\ \hline 
			Haro11 & 0h36m52.4544s & -33d33m16.7652s & 0.020598 & Dwarf & - & - &  ... & 10.67 & 37.02 & COR15 & HOW10 \\
			IRAS00397-1312 & 0h42m15.5119s & -12d56m3.3108s & 0.261717 & ULIRG & - & - &  ... & 10.7\ensuremath{\pm}0.2 & 369.5 & VEI09,PS17 & VIK17 \\
			NGC253 & 0h47m33.0727s & -25d17m18.996s & 0.000811 & HII & - & - &  ... & 10.4\ensuremath{\pm}0.1 & 4.7\ensuremath{\pm}0.9 & B-S09 & PAR18 \\
			HS0052+2536 & 0h54m56.3647s & +25d53m8.0052s & 0.045385 & Dwarf & - & - &  ... & 9.09\ensuremath{\pm}0.06 & 2.5\ensuremath{\pm}0.3 & COR15 & VIK17 \\
			UM311 & 1h15m34.403s & -0d51m46.06s & 0.005586 & Dwarf & - & - &  ... & 6.69\ensuremath{\pm}0.07 & 0.040\ensuremath{\pm}0.012 & COR15 & GalDR8 \\
			NGC625 & 1h35m5.1598s & -41d26m8.808s & 0.001321 & Dwarf & - & - &  ... & 8.67\ensuremath{\pm}0.1 & 0.0562\ensuremath{\pm}0.007 & COR15 & PAR18 \\
		\end{tabular}
		\tablefoot{Column (1): name of the galaxy. Columns (2) and (3): coordinates. Column (4): redshift. Column (5): spectral type. Columns (6)-(19): IR emission line fluxes and their errors in 1e$^{-14}$ erg/s/cm$^{2}$. Column (20): stellar masses. Column (21): star formation rates (SFR). Column (22): references for IR line fluxes. Column (23): references for stellar properties. The complete version of this table is available at the CDS.}
		\tablebib{IR fluxes. ARM07 \citep{Armus_2007}, B-S09 \citep{Bernard-Salas_2009}, BRE19 \citep{Breuck_2019}, COR15 \citep{Cormier_2015}, DAN05 \citep{Dannerbauer_2005}, FAR07 \citep{Farrah_2007}, FER15 \citep{Ferkinhoff_2015}, FO16 \citep{Fernandez-Ontiveros_2016}, FO21 \citep{Fernandez-Ontiveros_2021}, G+A09 \citep{Goulding_2009}, HC18 \citep{Herrera-Camus_2018}, IMA10 \citep{Imanishi_2010}, INA13 \citep{Inami_2013}, LAM18 \citep{Lamarche_2018}, NOV19 \citep{Novak_2019}, PS17 \citep{Pereira-Santaella_2017}, RIG18 \citep{Rigopoulou_2018}, TAD19 \citep{Tadaki_2019}, UZG16 \citep{Uzgil_2016}, VEI09 \citep{Veilleux_2009}.
		}
	\tablebib{Stellar properties. GalDR8 Galspec Data Release 8. HOW10 \citep{Howell_2010}, PAR18 \citep{Parkash_2018}, SHE10 \citep{Sheth_2010}, VIK17 \citep{Vika_2017}.
	}
	\end{table}

	\begin{table}
		\caption{List of IR fluxes for our sample of AGN.}\label{TabA2}
		\centering
		\begin{tabular}{llllllllllll}
			\textbf{Name} & \textbf{RA} & \textbf{De} & \textbf{z} & \textbf{Type} & \boldmath$\mathrm{Br}_{\alpha }$ & \textbf{[\ion{Ar}{ii}]7$\mu$ m} & ... & \textbf{[\ion{O}{iii}]88$\mu$ m} & \textbf{[\ion{N}{ii}]122$\mu$ m} & \textbf{[\ion{N}{ii}]205$\mu$ m} & \textbf{Ref.} \\  
			\textbf{(1)} & \textbf{(2)} & \textbf{(3)} & \textbf{(4)} & \textbf{(5)} & \textbf{(6)} & \textbf{(7)} & ... & \textbf{(21)} &  \textbf{(22)} & \textbf{(23)} & \textbf{(24)} \\ \hline 
			IRAS00198-7926 & 00h21m53.6141s & -79d10m07.9572s & 0.0728 & S2 & - & - & ... & 12.51\ensuremath{\pm}4.5 & - & - & FO16,PD22 \\
			NGC185 & 00h38m57.8837s & +48d20m14.6616s & -0.000674 & S2 & - & - & ... & - & - & - & FO16,PD22 \\
			MCG-01-24-012 & 09h20m46.2653s & -08d03m21.9564s & 0.019644 & S2 & - & - & ... & - & - & - & FO16,PD22 \\
			NGC4593 & 12h39m39.4550s & -05d20m39.0156s & 0.0091 & S1.0 & - & - & ... & 4.09\ensuremath{\pm}0.48 & 2.11\ensuremath{\pm}0.25 & - & FO16,PD22 \\
			NGC5506 & 14h13m14.8757s & -03d12m27.6984s & 0.006181 & S1h & - & - & ... & 102.26\ensuremath{\pm}3.31 & 14.14\ensuremath{\pm}1.15 & - & FO16,PD22 \\
			Mrk1383 & 14h29m06.5710s & +01d17m06.2196s & 0.08657 & S1.0 & - & - & ... & - & - & - & FO16,PD22 \\
		\end{tabular}
		\tablefoot{Column (1): name of the galaxy. Columns (2) and (3): coordinates. Column (4): redshift. Column (5): spectral type. Columns (6)-(23): IR emission line fluxes and their errors in 1e$^{-14}$ erg/s/cm$^{2}$. Column (24): references for IR line fluxes. The complete version of this table is available at the CDS.}
		\tablebib{ALO00 \citep{Alonso-Herrero_2000}, ARM07 \citep{Armus_2007}, B-S09 \citep{Bernard-Salas_2009}, BEL03 \citep{Bellamy_2003}, BEL04 \citep{Bellamy_2004}, BEN04 \citep{Bendo_2004}, BRA08 \citep{Brahuer_2008}, DAN05 \citep{Dannerbauer_2005}, FO16 \citep{Fernandez-Ontiveros_2016}, GOL95 \citep{Goldader_1995}, HC16 \citep{Hernan-Caballero_2016}, HC18 \citep{Herrera-Camus_2018}, IMA04 \citep{Imanishi_2004}, IMA10 \citep{Imanishi_2010}, INA13 \citep{Inami_2013}, INA18 \citep{Inami_2018}, KIM15 \citep{Kim_2015}, LAM17 \citep{Lamperti_2017}, LAN96 \citep{Lancon_1996}, LUT02 \citep{Lutz_2002}, MAR10 \citep{Martins_2010}, MUE11 \citep{Muller_2011}, MUR01 \citep{Murphy_2001}, PD22 \citep{Perez-Diaz_2022}, PEN21 \citep{Peng_2021}, PIQ12 \citep{Piqueras_2012}, PS17 \citep{Pereira-Santaella_2017}, REU02 \citep{Reunanen_2002}, REU03 \citep{Reunanen_2003}, RIF06 \citep{Riffel_2006}, SEV01 \citep{Severgnini_2001}, SMA12 \citep{Smajic_2012}, SPI21 \citep{Spinoglio_2022}, SPO22 \citep{Spoon_2022}, VEI97 \citep{Veilleux_1997}, VEI09 \citep{Veilleux_2009}, YAN21 \citep{Yano_2021}.
		}
	\end{table}
\end{landscape}

\begin{table*}
	\caption{Chemical abundances estimated from \textsc{HCm-IR}, using the grid of POPSTAR for our sample of SFGs.}\label{TabA3}
	\centering
	\begin{tabular}{lllll}
		\textbf{Name} & \boldmath$12+\log(O/H)$ & \boldmath$12+\log(S/H)$ & \boldmath$\log(N/O)$ & \boldmath$\log(U)$\\ \textbf{(1)} & \textbf{(2)} & \textbf{(3)} & \textbf{(4)} & \textbf{(5)} \\ \hline 
		Haro11 & 8.15\ensuremath{\pm}0.09 & 6.39\ensuremath{\pm}0.16 & -1.07\ensuremath{\pm}0.07 & -2.62\ensuremath{\pm}0.08 \\
		IRAS00397-1312 & 8.42\ensuremath{\pm}0.19 & 6.79\ensuremath{\pm}0.11 & - & -2.93\ensuremath{\pm}0.13 \\
		NGC253 & 8.72\ensuremath{\pm}0.23 & - & -0.92\ensuremath{\pm}0.25 & -3.27\ensuremath{\pm}0.22 \\
		HS0052+2536 & 8.08\ensuremath{\pm}0.22 & 6.29\ensuremath{\pm}0.29 & - & -2.65\ensuremath{\pm}0.12 \\
		UM311 & 8.12\ensuremath{\pm}0.18 & 6.51\ensuremath{\pm}0.19 & - & -2.68\ensuremath{\pm}0.09 \\
		NGC625 & 8.03\ensuremath{\pm}0.16 & 6.4\ensuremath{\pm}0.2 & - & -2.53\ensuremath{\pm}0.1 \\
		
	\end{tabular}
	\tablefoot{Column (1): name of the galaxy. Columns (2)-(5): chemical abundances and ionization parameters with their corresponding uncertainties. The complete version of this table is available at the CDS.}
\end{table*}

\begin{table*}
	\caption{Chemical abundances estimated from \textsc{HCm-IR}, using the grid of AGN models for $\alpha_{OX} = 0.8$ and the stopping criteria of $2\%$ of free electrons for our sample of AGNs.}\label{TabA4}
	\centering
	\begin{tabular}{lllll}
		\textbf{Name} & \boldmath$12+\log(O/H)$ & \boldmath$12+\log(S/H)$ & \boldmath$\log(N/O)$ & \boldmath$\log(U)$\\ \textbf{(1)} & \textbf{(2)} & \textbf{(3)} & \textbf{(4)} & \textbf{(5)} \\ \hline
		IRAS00198-7926 & 8.18\ensuremath{\pm}0.34 & 6.76\ensuremath{\pm}0.35 & - & -1.81\ensuremath{\pm}0.33 \\
		NGC185 & - & - & - & - \\
		MCG-01-24-012 & 7.86\ensuremath{\pm}0.36 & 6.35\ensuremath{\pm}0.45 & - & -1.6\ensuremath{\pm}0.42 \\
		NGC4593 & 8.07\ensuremath{\pm}0.36 & 6.41\ensuremath{\pm}0.51 & -0.91\ensuremath{\pm}0.22 & -1.78\ensuremath{\pm}0.34 \\
		NGC5506 & 8.15\ensuremath{\pm}0.3 & 6.68\ensuremath{\pm}0.39 & -0.74\ensuremath{\pm}0.15 & -1.88\ensuremath{\pm}0.37 \\
		Mrk1383 & 7.83\ensuremath{\pm}0.38 & 6.2\ensuremath{\pm}0.47 & - & -1.58\ensuremath{\pm}0.4 \\
		
	\end{tabular}
	\tablefoot{Column (1): name of the galaxy. Columns (2)-(5): chemical abundances and ionization parameters with their corresponding uncertainties. The complete version of this table is available at the CDS.}
\end{table*}

\end{appendix}

\end{document}